\documentclass[runningheads,a4paper]{llncs}

\usepackage{alltt}
\usepackage{import}
\usepackage{multirow}
\usepackage{microtype}
\usepackage{array}
\usepackage{graphicx,wrapfig}
\usepackage[noadjust]{cite}
\usepackage{tikz}
\usetikzlibrary{plotmarks}
\usepackage{pgfplotstable}
\usepackage{filecontents}
\usepackage{pgfplots}
\usepackage{amsfonts}
\usepackage{amssymb}
\usepackage{amsmath}
\usepackage{stmaryrd}
\usepackage{eucal}
\usepackage{color}
\usepackage{listings}
\usepackage{verbatim}
\usepackage{psfrag}
\usepackage{epsfig}
\usepackage{wasysym} 
\usepackage{paralist}
\usepackage[algo2e,linesnumbered,ruled,lined]{algorithm2e}
\usepackage{hyperref}
\usepackage[subnum]{cases}
\makeatother

\newcommand{\extendedonly}[1]{}

\newcommand{\Omit}[1]{}

\newcommand{\rmcmt}[1]{{\color{magenta}{#1}}}

\newcommand{\sat}{\textsc{sat}\xspace}

\renewcommand{\vec}[1]{{\boldsymbol #1}}
\newcommand{\vecv}[2]{{\left(\begin{array}{@{}c@{}} #1 \\ #2\end{array}\right)}}
\newcommand{\qmat}[4]{{\left(\begin{array}{@{}cc@{}} #1 & #2 \\  #3 & #4\end{array}\right)}}
\newcommand{\mat}[1]{{\boldsymbol #1}}

\newcommand{\true}{\mathsf{true}}
\newcommand{\false}{\mathsf{false}}

\newcommand{\meet}{\ensuremath{\sqcap}}

\newcommand{\program}{\mathit{Prog}}
\newcommand{\assertions}{\mathit{Assn}}
\newcommand{\assertion}{\mathit{a}}
\newcommand{\constraints}{\Sigma}
\newcommand{\constraint}{\sigma}
\newcommand{\formula}{\varphi}

\newcommand{\vars}{\mathit{Vars}}
\newcommand{\subvars}{\mathit{V}}
\newcommand{\boolvars}{\mathit{BVars}}
\newcommand{\numvars}{\mathit{NVars}}

\newcommand{\booldomain}{\mathcal{B}}
\newcommand{\reldomain}{\mathcal{TP}}

\newcommand{\numvar}{x}
\newcommand{\numconcval}{\mathit{X}}
\newcommand{\numabsval}{d}
\newcommand{\decisionvar}{\mathit{q}}
\newcommand{\conflictset}{\mathit{C}}

\newcommand{\abstrans}[2]{\llbracket #2 \rrbracket_{#1}}

\newcommand{\abd}[2]{abd_{#1}^{#2}}
\newcommand{\conf}[2]{conf_{#1}^{#2}}

\newcommand{\abstransset}{\mathcal{A}}
\newcommand{\abstransel}[1]{\mathit{ded}^{#1}}
\newcommand{\domain}{\mathit{D}}
\newcommand{\subdomain}{\mathit{L}}

\newcommand{\val}{\mathit{u}}

\newcommand{\absval}{\mathit{a}}

\newcommand{\newdeductions}{\mathit{v}}
\newcommand{\intervals}{\mathit{Itvs}}
\newcommand{\octagons}{\mathit{Octs}}
\newcommand{\onlynew}{\mathit{onlyNew}}
\newcommand{\aunit}{\mathit{AUnit}}

\newcommand{\abs}{\mathit{abs}}
\newcommand{\trail}{\mathcal{T}}
\newcommand{\reasons}{\mathcal{R}}
\newcommand{\worklist}{\mathit{worklist}}
\newcommand{\initworklist}{\mathit{initWorklist}}
\newcommand{\updateworklist}{\mathit{updateWorklist}}
\newcommand{\makesubdomain}{\mathit{MakeL}}

\newcommand{\decide}{\mathit{decide}}
\newcommand{\deduce}{\mathit{deduce}}
\newcommand{\analyzeconflict}{\mathit{analyzeConflict}}

\newcommand{\propheur}{\mathit{H_P}}
\newcommand{\decheur}{\mathit{H_D}}
\newcommand{\confheur}{\mathit{H_C}}

\newcommand{\safe}{\ensuremath{\mathit{Safe}}}

 \SetKwSty{fpfont}
 \SetFuncSty{fpfn}

 \SetKwFunction{Decide}{decide}
 \SetKwFunction{TransProp}{tprop}
 \SetKwFunction{UnitProp}{uprop}
 \SetKwFunction{Refines}{refines}
 \SetKwFunction{Deduce}{deduce}
 \SetKwFunction{IsUnit}{is\_unit}
 \SetKwFunction{Propagate}{propagate}
 \SetKwFunction{Invariant}{invariant}
 \SetKwFunction{Atomic}{atomic}
 \SetKwFunction{Learn}{learn}
 \SetKwFunction{Backtrack}{backtrack}
 \SetKwFunction{Clausegen}{clauseGen}
 \SetKwFunction{Invariantgen}{invariantGen}

 \SetKw{Let}{let}

\SetKwFunction{Asserting}{asserting}
\SetKwFunction{Cutheur}{cutheur}
\SetKwFunction{Generalise}{generalise}
\SetKwFunction{Analyse}{analyse}
\SetKwFunction{ClauseGeneralise}{clauseGeneralise}
\SetKwFunction{WpGeneralise}{wpGeneralise}
\SetKw{MyCase}{case}
\SetKwFunction{SetReason}{setReason}
\SetKwFunction{PickAssertingCut}{assertingCut}

\newcommand{\decomp}{{\mathit{decomp}}}

\renewcommand{\min}{\mathit{min}}

\begin{document}

\title{Lifting CDCL to Template-based Abstract Domains
for Program Verification \\ (Extended Version)
}

\author{Rajdeep Mukherjee\inst{1} \and Peter Schrammel\inst{2} \and 
Leopold Haller\inst{3} \and \\ 
Daniel Kroening\inst{1} \and Tom Melham\inst{1}}

\authorrunning{Mukherjee, Schrammel, Haller, Kroening, Melham}

\institute{University of Oxford, UK \and University of Sussex, UK \and Google Inc., USA}

\maketitle

\begin{abstract}
%
The success of Conflict Driven Clause Learning (CDCL) for Boolean
satisfiability has inspired adoption in other domains.  We~present a novel
lifting of CDCL to program analysis called \emph{Abstract Conflict Driven
Learning for Programs} (ACDLP).  ACDLP alternates between \emph{model
search}, which performs over-approximate deduction with constraint
propagation, and \emph{conflict analysis}, which performs under-approximate
abduction with heuristic choice.  We~instantiate the model search and
conflict analysis algorithms to an abstract domain of \textit{template
polyhedra}, strictly generalizing CDCL from the Boolean lattice to a richer
lattice structure.  Our template polyhedra can express intervals, octagons
and restricted polyhedral constraints over program variables.  We have
implemented ACDLP for automatic bounded safety verification of C programs. 
We~evaluate the performance of our analyser by comparing with CBMC, which
uses CDCL, and Astr{\'e}e, a commercial abstract interpretation tool. 
We~observe two orders of magnitude reduction in the number of decisions,
propagations, and conflicts as well as a~1.5x speedup in runtime compared to
CBMC.  Compared to Astr{\'e}e, ACDLP solves twice as many benchmarks and has
much higher precision.  This is the first instantiation of CDCL with a
template polyhedra abstract domain.
\end{abstract}

\section{Introduction}

%
Static program analysis with abstract interpretation~\cite{CC77} is widely
used to verify properties of safety-critical systems.  Static analyses
commonly aim to compute program invariants as fixed-points of abstract
transformers.  Abstract states are chosen from a lattice that has meet
$(\sqcap)$ and join $(\sqcup)$ operations; the meet precisely models set
intersection (or conjunction, taking a logical view), and the join
over-approximates set union (or disjunction).  Over-approximation in the
join operation is one of the sources of precision loss, which causes false
alarms.  Typical abstract domains are non-distributive; suppose $a$ and $b$
together represent the abstract semantics of a program and $c$ represents a
set of abstract behaviours that violate the specification.  In a
non-distributive domain, $(a \sqcup b) \sqcap c$ can be strictly less
precise than $(a \sqcap c) \sqcup (b \sqcap c)$.  This means that in typical
abstract domains, analysing program behaviours separately can improve the
precision of the analysis.  Usual means to address false alarms therefore
include not only the use of richer abstract domains, but also of refinements
that delay joins or perform some form of case-splitting.  Such techniques
trade off higher precision against lower efficiency and may be susceptible
to case enumeration behaviour.

By contrast, Model Checking (MC)~\cite{mc-book} can be seen to operate on
distributive lattice structures that represent disjunction without loss
of precision.  Classical MC directly operates on distributive
representations, such as BDDs, while more recent implementations use SAT
solvers.  SAT solvers themselves operate on partial assignments, which are
non-distributive structures.  To handle disjunction, case-splitting is
performed~\cite{sas12}.  Propositional SAT solvers solve large formulae, and
are often able to avoid enumerating cases.  The impressive performance of
modern solvers is credited to well-tuned decision heuristics and
sophisticated clause learning algorithms.  Collectively, these algorithms
are referred to as \emph{Conflict Driven Clause Learning}
(CDCL)~\cite{cdcl}.  An obvious idea is to lift CDCL from the domain of 
partial assignments to other non-distributive domains.

Abstract Conflict Driven Clause Learning (ACDCL)~\cite{dhk2013-popl} is one
such lattice-based generalization of CDCL.  ACDCL is a general algorithmic
framework, parameterized by a concrete domain~$C$ and an abstract
domain~$A$.  Classical CDCL can be viewed as an instance of ACDCL in which
$C$ is the set of propositional truth assignments and $A$ the domain of
propositional partial assignments~\cite{leo-thesis}.  Since the concrete
domain of interest is a parameter to the framework, ACDCL can in principle
be used to build both \emph{logical decision
procedures}~\cite{DBLP:journals/fmsd/BrainDGHK14} and \emph{program
analyzers}.  In the former case, the concrete domain is the set of candidate
models for the formula; in the latter case, it is the set of program traces
that may lead to an error.  Haller
et~al.~in~\cite{DBLP:journals/fmsd/BrainDGHK14} illustrate the first idea by
presenting a floating-point decision procedure that uses interval constraint
propagation.


In this paper, we explore the second idea by presenting an extension of 
ACDCL to program analysis.  We call our 
framework \emph{Abstract Conflict Driven Learning for Programs}
(ACDLP).  The key insight of ACDLP is to use decisions and learning to precisely 
reason about disjunctions in non-distributive domains, thereby automatically 
refining the precision of analysis for safety checking of C programs.  
We~introduce two central components of our framework: an abstract
model search algorithm that uses decisions and propagations to 
search for counterexample trace and an abstract conflict analysis 
procedure that approximates a set of unsafe traces through transformer 
learning.
We illustrate the application of our framework to program analysis 
using a \textit{template polyhedra abstract domain}~\cite{vmcai05}, 
which includes most of the commonly used abstract domains, such as boxes, 
octagons, zones and TCMs.  

We present an experimental evaluation of our analyser compared 
to CBMC~\cite{cbmc.tacas:2004}, which uses propositional solvers, and to 
Astr{\'e}e~\cite{DBLP:conf/pldi/BlanchetCCFMMMR03}, a commercial abstract 
interpretation tool.  In this paper, we make the following contributions.

\Omit{
Our experiments suggest that ACDLP can be seen as a 
technique to improve efficiency of SAT-based BMC. Additionally, it can 
also be perceived as an automatic way to improve the precision of abstract 
interpretation for bounded unwindings of programs.  
}
\begin{enumerate}
\item A novel program analysis framework that lifts model search and conflict
analysis procedures of CDCL algorithm over a template polyhedra abstract domain. These techniques
are embodied in our tool, \emph{ACDLP}, for automatic bounded safety verification 
    of C programs.
\item A parameterized abstract transformer that guides the model
  search in forward, backward and multi-way direction for counterexample
    detection. 
\item A conflict analysis procedure that performs UIP-based transformer learning 
  over template polyhedra abstract domain through abductive reasoning. 
\end{enumerate}
\section{Motivating Examples}

\begin{figure}[t]
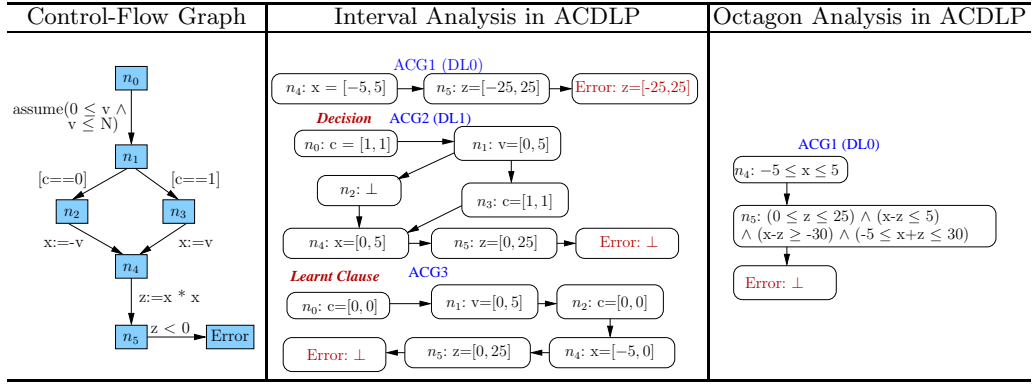

\centering
\begin{tabular}{c|c|c}
\hline
Control-Flow Graph & Interval Analysis in ACDLP & Octagon Analysis in ACDLP\\
\hline
\scriptsize
\begin{minipage}{3.3cm}
\scalebox{.52}{\import{figures/}{example.pspdftex}}
\end{minipage}
&
\begin{minipage}{5.7cm}
\vspace*{0.3cm}
\scalebox{.5}{\import{figures/}{acdl_run.pspdftex}}\vspace*{0.1cm}
\end{minipage}
&
\begin{minipage}{3.6cm}
\vspace*{0.3cm}
\scalebox{.5}{\import{figures/}{acdl_oct.pspdftex}}
\end{minipage}
\\
\hline
\end{tabular}
\caption{\label{fig:example1}
CFG and corresponding Abstract Conflict Graphs for Interval and Octagon Analysis}
\end{figure}

In this section, we present two simple examples to demonstrate the core idea
of ACDLP for bounded verification.  For each example, we apply three
analysis techniques: \textit{abstract interpretation} (AI), SAT-based
\textit{bounded model checking} (BMC) and ACDLP. \\ 
\textbf{First Example} The simple Control-Flow Graph (CFG) in
Fig.~\ref{fig:example1} squares a machine integer and checks that the result
is positive.  To avoid overflow, we assume the input~v has an upper bound~N. 
This example demonstrates that a) interval analysis in ACDLP is more precise
than a forward AI in the interval domain, and b) ACDLP with intervals can
achieve a precision similar to that of AI with octagons without employing
more sophisticated mechanisms such as trace partitioning~\cite{toplas07}.

\noindent \textbf{AI versus ACDLP}
Conventional forward interval AI is too imprecise to verify safety of this
program owing to the control-flow join at node~$n_4$.  For example, the
state-of-the-art AI tool Astr{\'e}e requires external hints, provided by
manually annotating the code with partition directives at $n_1$.  This
tells Astr{\'e}e to analyse the program paths separately.

However, ACDLP can be understood as an algorithm to automatically infer such
partitions.  For the example in Fig.~\ref{fig:example1}, interval analysis
with ACDLP is sufficient to prove safety.  The analysis records the
decisions and deductions in a \textit{trail} data-structure.  The trail can
be viewed to represent a graph structure called the {\em Abstract Conflict
Graph} (ACG) that stores dependencies between decisions and deductions
nodes, similar to the way an \emph{Implication Graph}~\cite{cdcl} works in a SAT solver. 
Nodes of the ACG in the second column of Fig.~\ref{fig:example1} are
labelled with the CFG location and the corresponding abstract value. 
Beginning with the assumption that v=[0,5] at node~$n_1$, the intervals
generated by forward analysis in the initial deduction phase at
\emph{decision level}~0 (DL0) are $x=[-5,5]$ and $z=[-25,25]$.  Clearly
these do not prove safety, as shown in ACG1.  ACDLP therefore makes a
heuristic decision, at DL1, to refine the analysis.  With the decision
$c=[1,1]$, interval analysis then concludes $x=[0,5]$ at node~$n_4$, which
leads to $(\mathrm{Error}{:}\;\bot)$ in ACG2, indicating that the error
location is unreachable and that the program is safe when $c=[1,1]$.

Reaching $(\mathrm{Error}{:}\;\bot)$ is analogous to reaching a conflict in
a propositional SAT solver.  At this point, a clause-learning SAT solver
learns a reason for the conflict and backtracks to a level such that the
learnt clause is \textit{unit}.  By a similar process, ACDLP learns that
$c=[0,0]$.  That is, all error traces must satisfy $(c \neq 1)$.  The
analysis discards all interval constraints that lead to the conflict and
backtracks to DL0.  ACDLP then performs interval analysis with the learnt
clause $(c \neq 1)$.  This also leads to a conflict, as shown in ACG3.  The
analysis cannot backtrack further and so terminates, proving the program
safe.  Thus, {\em decision} and {\em clause learning} are used to infer the
partitions necessary for a precise analysis.  Alternatively, the octagon
analysis in ACDLP---illustrated in the third column of
Fig.~\ref{fig:example1}---can prove safety with propagations only. 
No~decisions are required.  Forward AI with octagons in Astr{\'e}e is also
able to prove safety.  
\begin{table}[!b]
\begin{center}
{
\begin{tabular}{l|r|r|r|r|r|r}
\hline
  Solver & Domains & decisions & propagations & conflicts & conflict literals & restarts \\ \hline
  \multicolumn{7}{c}{Solver statistics for Fig.~\ref{fig:example1} (For N = 46000)} \\ \hline
  MiniSAT & $\boolvars \rightarrow \{t,f,?\} $ & 233 & 36436 & 162 & 2604 & 2 \\ \hline
  ACDLP & $\mathit{nodes} \rightarrow \intervals[\numvars]$ & 1 & 17 & 1 & 1 & 0 \\ \hline
  ACDLP & $\mathit{nodes} \rightarrow \octagons[\numvars]$ & 0 & 7 & 0 & 0 & 0 \\ \hline 
  \multicolumn{7}{c}{Solver statistics for Fig.~\ref{fig:example2}} \\ \hline
  MiniSAT & $\boolvars \rightarrow \{t,f,?\} $ & 4844 & 32414 & 570 & 4750 & 5 \\ \hline
  ACDLP & $\mathit{nodes} \rightarrow \octagons[\numvars]$ & 4 & 412 & 2 & 2 & 0 \\ 
\hline
\end{tabular}
}
\end{center}
\caption{SAT-based BMC versus ACDLP for verification of programs in Fig.~\ref{fig:example1} and
  Fig.~\ref{fig:example2}}
\label{solver}
\end{table}
\begin{figure}[t]
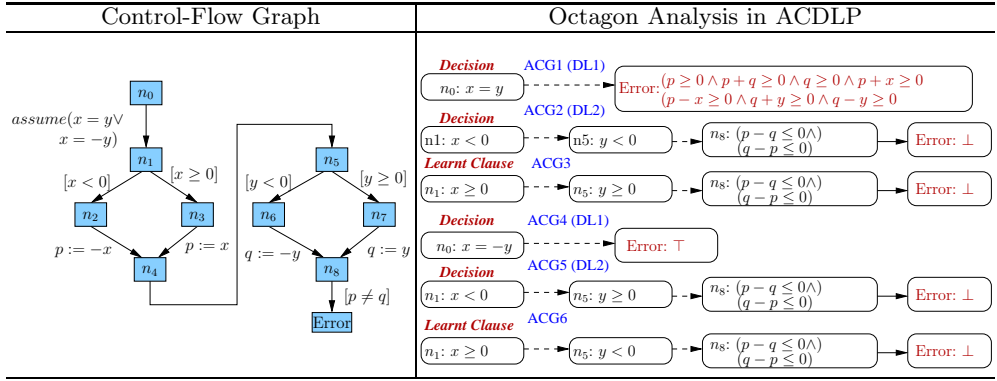

\centering
\begin{tabular}{c|c}
\hline
Control-Flow Graph & Octagon Analysis in ACDLP \\
\hline
\scriptsize
\begin{minipage}{5.28cm}
\scalebox{.52}{\import{figures/}{cfg.pspdftex}}
\end{minipage}
&
\begin{minipage}{7.5cm}
\vspace*{0.3cm}
\scalebox{.5}{\import{figures/}{oct_partition.pspdftex}}\vspace*{0.1cm}
\end{minipage}
\\
\hline
\end{tabular}\caption{\label{fig:example2}
CFG and corresponding Abstract Conflict Graphs for Octagon Analysis}
\end{figure}

\noindent \textbf{Second Example} Fig.~\ref{fig:example2}, demonstrates that 
octagon analysis in ACDLP is more precise than forward AI in the octagon domain.
The CFG in Fig.~\ref{fig:example2} computes the absolute 
values of two variables, $x$ and~$y$, under the assumption 
$(x=y) \lor (x=-y)$. 

\noindent \textbf{AI versus ACDLP}
Forward AI in the octagon domain infers 
the octagonal constraint $\mathrm{Error}{:}\;(p\geq0 \wedge p+q\geq0 \wedge q\geq0 \wedge p+x\geq0 \wedge 
p-x\geq0 \wedge q+y\geq0 \wedge q-y\geq0)$. Clearly this is too 
imprecise to prove safety. 
The octagonal analysis in ACDLP is illustrated by the ACGs in
Fig.~\ref{fig:example2}.  Due to space limitations, we elide intermediate
deductions with dotted lines.  The decision $x=y$ at DL1 is not sufficient
to prove safety, as shown in ACG1.  Hence, a new decision $x<0$ is made at
DL2, followed by forward propagation that infers $y<0$ at node~$n_5$.  This
subsequently leads to safety ($\mathrm{Error}{:}\;\bot$), as shown in ACG2. 
The analysis learns the reason for the conflict, discards all deductions in
ACG2 and backtracks to DL1.  Octagon analysis is run with the learnt
constraint $(x\geq0)$ and this infers $y\geq0$ at node~$n_5$, as shown in ACG3. 
This also leads to safety ($\mathrm{Error}{:}\;\bot$).  The analysis now
makes a new decision $x=-y$ at DL1.  The procedure is repeated leading to
results shown in ACG4, ACG5, and ACG6.  Clearly, the decisions $x=-y$ and
$x<0$ also lead to safety.  The analysis backtracks to DL0 and returns {\em
safe}.  Note that the specific decision heuristic we use in this case
exploits the control structure of the program to infer partitions that are
sufficient to prove safety. \\ \\
\noindent \textbf{ACDLP versus BMC}
ACDLP can require many fewer iterations than SAT-based BMC due to its
ability to reason over much richer lattice structures.
A~SAT-based BMC converts the program into a bit-vector equation 
and passes that to a CDCL-based SAT solver for proving safety.  
Table~\ref{solver} compares the statistics for BMC with 
MiniSAT~\cite{minisat} solver to those for interval and octagon 
analysis in ACDLP. In the column labelled $\textrm{Domains}$, 
$\boolvars$ is the set of propositional variables; each of these is mapped to
{\em true} (t), {\em false} (f) or {\em unknown} $(?)$. $\numvars$ is the set of
numerical variables, $\mathit{nodes}$ the set of nodes in the CFG;
$\intervals[\numvars]$ and $\octagons[\numvars]$ are the Interval and Octagon
domains over $\numvars$. As can be seen, ACDLP outperforms BMC in the total number of 
{\em decisions}, {\em propagations}, {\em learnt clauses} and {\em restarts} 
for both example programs.

\Omit{
However, in contrast to a SAT solver, the octagonal analysis in ACDLP is 
very efficient, as shown in Table~\ref{solver2}.  \rmcmt{This shows the advantage 
of using richer abstract domains within the CDCL-style analysis}.  
\begin{table}[!b]
\begin{center}
{
\begin{tabular}{l|r|r|r|r|r|r}
\hline
  Solver & Domain & decisions & propagations & conflicts & conflict literals & restarts \\ \hline
  MiniSAT & $\boolvars \rightarrow \{t,f,?\} $ & 4844 & 32414 & 570 & 4750 & 5 \\ \hline
  ACDLP & $\mathit{nodes} \rightarrow $ & 4 & 412 & 2 & 2 & 0 \\ 
\hline
\end{tabular}
}
\end{center}
\caption{Solver statistics for Example given in Fig.~\ref{fig:example2}}
\label{solver2}
\end{table}
}


\section{Program Model and Abstract Domain}\label{sec:domains}
\subsection{Program Representation}\label{sec:program}  
We consider \emph{bounded programs} with safety
properties given as a set of assertions, $\assertions$, in the program.
A bounded program is obtained by a transformation that unfolds
loops and recursions a finite number of times. The result 
is represented by a set
$\constraints=\program\cup\{\neg \bigwedge_{\assertion\in\assertions} \assertion\}$,
where $\program$ contains an encoding of the statements in the program as
constraints, obtained after translating the program
into single static assignment (SSA) form via a data flow analysis.
The representation $\constraints$ for the program in Fig.~\ref{fig:example1} is 
\begin{equation}\label{eq:ssa}
\begin{array}{l@{}l}
\{ & g_0 = (0 \leq v \leq N),\:
     g_1 = (g_0 \wedge c),\:
     x_0 = v,\:
     x_1 = -v, \\ 
  &  x_2 = g_1?x_0:x_1, \:
     g_2 = (g_1 \vee g_0\wedge \neg c), \:
     z  = x_2{\cdot}x_2,\:
     g_2 \wedge z{<}0 \}
\end{array}
\end{equation}
Assignments such as x:=v become equalities $x_1=v$, where the
left-hand side variable gets a subscripted fresh name.
Control flow is encoded using guard variables, e.g.~$g_1=g_0\wedge c$.
Data flow joins become conditional expressions, e.g.~$x_3=g_1?x_1:x_2$.
The assertions in $\assertions$ are constraints such as $g_2
\Rightarrow z\geq 0$, meaning that if $g_2$ holds
(i.e.,~the assertion is reachable) then the assertion must hold.
We~write $\vars$ for the set of variables occurring in $\constraints$.  
Based on the above program representation, we define a \textit{safety formula}
($\formula$) as the conjunction of everything in $\constraints$, that is,  
$\formula:= \bigwedge_{\constraint\in\constraints} \constraint$. The formula 
$\formula$ is unsatisfiable if and only if the program is safe.
\begin{table}[t]
\small
\begin{center}
{
\begin{tabular}{c@{\quad}|@{\quad}c@{\quad}|@{\quad}c@{\quad}|@{\quad}c@{\quad}|@{\quad}c}
Interval & Octagons & Zones & Equality & Fixed-coef.~Polyhedra \\ \hline
$a \leq x_i \leq b$ & $\pm x_i \pm x_j \leq d$ & $x_i - x_j \leq d$ & 
  $x_i=x_j\Omit{,(x_i \neq x_j)}$ & $a_1x_1 + \ldots + a_nx_n \leq d$
\end{tabular}
}
\end{center}
\caption{Template instances in the template polyhedra domain}
\label{domain}
\end{table}
%
\subsection{Abstract Domain}
In this paper, we instantiate ACDLP over a reduced product domain~\cite{CC79}
$\domain[\vars]=\booldomain^{|\boolvars|}\times\reldomain[\numvars]$ where
$\booldomain$ is the Boolean domain that permits abstract values
$\{\true,\false,\bot,\top\}$ over boolean variables $\boolvars$ in the
program, and $\reldomain$ is a \textit{template polyhedra}~\cite{vmcai05}
domain over the numerical (bitvector) variables $\numvars$.  Our template
polyhedra domain can express various relational and non-relational templates
over $\numvars$, as given in Table~\ref{domain}.
%
\subsubsection{Template Polyhedra Abstract Domain}
%
An abstract value of the template polyhedra domain~\cite{vmcai05}
represents a set $\numconcval$ of values of the vector $\vec{\numvar}$ 
of numerical (bitvector) variables $\numvars$ of their respective
data types. (Currently, signed and unsigned integers are supported.)
For example, in the program given by Eq.~(\ref{eq:ssa}), we have four 
numerical variables, written as the vector $\vec{\numvar} = (x_0,x_1,x_2,z)$.  
An abstract value is a constant vector $\vec{\numabsval}$ that represents 
sets of values for $\vec{\numvar}$ for which 
$\mat{C}\vec{\numvar}\leq\vec{\numabsval}$, for a fixed coefficient 
matrix $\mat{C}$.  The domain containing $\vec{\numabsval}$ is augmented 
by a special element $\bot$ to denote the minimal element of the lattice.  
There are several optimisation-based 
techniques~\cite{vmcai05} for computing the domain operations, 
such as meet ($\meet$) and join ($\sqcup$), in the template polyhedra domain.  
In our implementation, we use the strategy iteration approach of~\cite{BJKS15}.
The abstraction function is defined by $\alpha(\numconcval) = \min \{\vec{\numabsval}\mid
\mat{C}\vec{\numvar}\leq\vec{\numabsval}, \vec{\numvar}\in \numconcval\}$, where 
$\min$ is applied component-wise.  The concretisation $\gamma(\vec{\numabsval})$ is the set $\{\vec{\numvar}\mid
\mat{C}\vec{\numvar}\leq\vec{\numabsval}\}$ and $\gamma(\bot)=\emptyset$,
i.e., the empty polyhedron.

For notational convenience we will use conjunctions of linear
inequalities, for example $x_1\geq 0 \wedge x_1-z\leq 30$, to write the
abstract domain value $\vec{\numabsval}=\vecv{0}{30}$,
with $\mat{C}=\qmat{-1}{0}{1}{-1}$ and $\vec{\numvar}=\vecv{x_1}{z}$; 
$\true$ corresponds to abstract value $\top$ and $\false$ to 
abstract value $\bot$.
For a program with $N=|\numvars|$ variables, the template 
matrix $\mat{C}$ for the interval domain $\intervals[\numvars]$, 
has $2N$ rows. Hence, it generates at most $2N$ inequalities, one
for the upper and lower bounds of each variable.
For octagons $\octagons[\numvars]$, we have at most $2N^2$
inequalities, one for the upper and lower bounds of each variable and
sums and differences for each pair of variables. 
Unlike a non-relational domain, a relational domain such as octagons 
requires the computation of a \emph{closure} in order to obtain a normal 
form, necessary for precise domain operation. 
The closure computes all implied domain constraints.  
An example of a closure computation for octagonal inequalities is
$\mathit{closure}((x-y \leq 4) \wedge (y-z \leq 5))=((x-y \leq 4) \wedge (y-z
\leq 5) \wedge (x-z \leq 9))$.
For octagons, closure is the most critical and expensive operator; it has  
cubic complexity in the number of program variables.  We therefore compute 
closure lazily in template polyhedra domain in our abstract model search 
procedure, which is described in section~\ref{lazyclosure}. 
\subsubsection{Abstract Transformers}
An abstract transformer $\abstrans{\domain}{\constraint}$ transforms an
abstract value $\absval$ through a constraint $\constraint$; it
\emph{deduces} information from $\absval$ and $\constraint$.  The best
transformer is
\begin{equation}\label{eq:abstrans}
\abstrans{\domain}{\constraint}(\absval)=\absval\meet\alpha(\{\val\mid \val\in\gamma(\absval), \val\models \constraint\})
\end{equation}
 where we write 
$\val\models \constraint$ if the concrete value $\val$ satisfies the constraint $\constraint$.
Any abstract transformer that over-approximates the best abstract
transformer is a sound transformer and can be used in our algorithm.
For example, we can deduce $\abstrans{\domain}{x=2(y+z)}(\absval)=(0\leq y
\leq 2 \wedge\allowbreak 1 \leq y-z \leq 1\wedge\allowbreak -2\leq x\leq 6)$
for the abstract value $\absval=(0\leq y \leq 2 \wedge\allowbreak 1 \leq y-z
\leq 1)$.
We~denote the set of abstract transformers for a safety formula
$\formula$ using the abstract domain $\domain$ by
$\abstransset=\{\abstrans{\domain}{\constraint}\mid
\constraint\in\constraints\}$.
%

\subsection{Properties of Abstract Domains}
%
An important property of a clause-learning SAT solver is that each
non-singleton element of the partial assignment domain can be 
decomposed into a set of \textit{precisely complementable} singleton
elements~\cite{dhk2013-popl}.  This property of domain elements
is necessary to learn elements that help to guide the model search away from the
conflicting region of the search space.  
Most numerical abstract domains, such as intervals and octagons lack
complements in general, i.e., not every element in the domain has a
precise complement.  However, these domain elements can be represented
as intersections of half-spaces, each of which admits a precise complement.
We formalise this in the sequel.
\begin{definition} 
A \emph{meet irreducible} $m$ in a complete lattice 
structure $A$ is an element with the following property.
\begin{equation}
\forall m_1, m_2 \in A: m_1 \meet m_2 = m \implies (m = m_1 \lor m = m_2), m \neq \top  
\end{equation}
\end{definition}
The meet irreducibles in the Boolean domain $\booldomain$ 
for a variable $x$ are $x$ and $\neg x$. The meet 
irreducibles in the template polyhedra domain are all elements 
that concretise to half-spaces, i.e., they can be represented 
by a single inequality. For the interval domain, these are 
$x \leq d$ or $x \geq d$ for constants $d$.
\begin{definition}
A \emph{meet decomposition} $\decomp(\absval)$ of an abstract
element $\absval \in \domain$ is a set of meet irreducibles $M \subseteq \domain$ such that 
$\absval=\bigsqcap_{m\in M} m$.
\end{definition}

\noindent For polyhedra this intuitively means that each polyhedron can be
written as an intersection of half-spaces.
For example, the meet decomposition of the interval domain element
$decomp(2\leq x\leq 4 \wedge 3\leq y\leq 5)$ is
the set $\{x\geq 2, x\leq 4, y\geq 3, y\leq 5\}$.
\begin{definition} 
An element $\absval\in \domain$ is called \emph{precisely complementable}
iff there exists $\bar{\absval} \in \domain$ such that 
$\neg\gamma(\bar{\absval})=\gamma(\absval)$.
That is, there is an element whose complemented concretisation equals
the concretisation of $\absval$.
\end{definition}
The precise complementation property of a partial assignment lattice can
be generalised to other lattice structures. 
For example, the precise complement of a meet irreducible $(x \leq 2)$ in
the interval domain over integers is $(x \geq 3)$, or the precise complement
of the meet irreducible $(x+y \leq 1)$ in the octagon domain over integers
is $(x+y \geq 2)$.  Our domain implementation supports precise
complementation operation.  However, standard abstract interpretation does
not require a complementation operator.  Hence, abstract domain libraries,
such as APRON~\cite{apron}, do not provide it.  But it can be implemented
with the help of a meet decomposition as explained above.
%


\section{Abstract Conflict Driven Learning for Programs}

\begin{figure}[htbp]
\centering
\scalebox{.6}{\import{figures/}{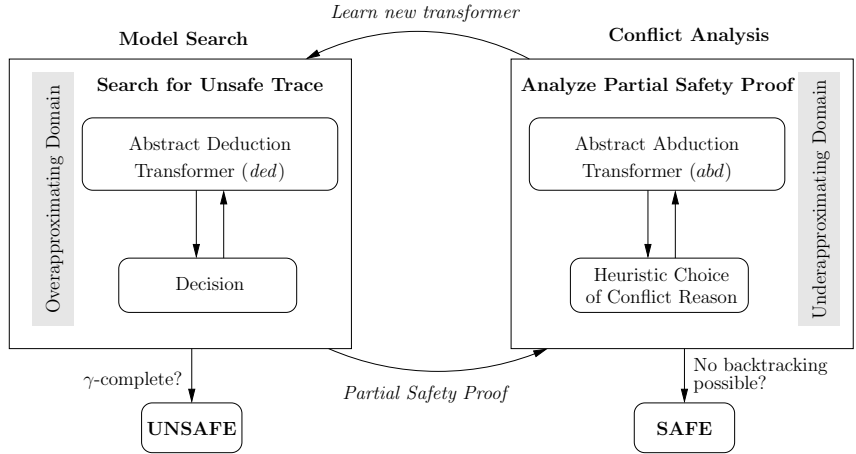}}
\caption{Architectural View of ACDLP\label{acdlp-top}}
\end{figure}

\begin{algorithm2e}[t]
\DontPrintSemicolon
\SetKw{return}{return}
\SetKwRepeat{Do}{do}{while}
\SetKwData{conflict}{conflict}
\SetKwData{safe}{safe}
\SetKwData{sat}{sat}
\SetKwData{unsafe}{unsafe}
\SetKwData{unknown}{unknown}
\SetKwData{true}{true}
\SetKwInOut{Input}{input}
\SetKwInOut{Output}{output}
\SetKwFor{Loop}{Loop}{}{}
\SetKw{KwNot}{not}
\begin{small}
\Input{A program in the form of a set of abstract transformers $\abstransset$.}
\Output{The status \safe or \unsafe. 
}
$\trail \leftarrow \langle\rangle$, $\reasons \leftarrow []$ \;
$\mathit{result} \leftarrow \deduce_{\propheur}(\abstransset,\trail,\reasons)$ \;
\lIf{$\mathit{result}$ = \conflict} {
  \return \safe}
\While{$true$} 
{
\lIf{$\mathit{result}$ = \sat} {
  \return \unsafe}
  $\decisionvar \leftarrow \decide_{\decheur}(\abs(\trail))$ \;
  $\trail \leftarrow \trail \cdot \decisionvar$ \; 
  $\reasons[|\trail|] \leftarrow \top$ \;
  $\mathit{result} \leftarrow \deduce_{\propheur}(\abstransset,\trail,\reasons)$\;
  \Do{$\mathit{result} = \conflict$} {
    \lIf{$\neg \analyzeconflict_{\confheur}(\abstransset,\trail,\reasons)$} {
      \return \safe
    }
    $\mathit{result} \leftarrow \deduce_{\propheur}(\abstransset,\trail,\reasons)$ \;
  }
}
\end{small}
\caption{Abstract Conflict Driven Learning $\mathit{ACDLP}_{\propheur,\decheur,\confheur}(\abstransset)$\label{Alg:acdcl}}
\end{algorithm2e}

Figure~\ref{acdlp-top} presents our framework called \emph{Abstract Conflict
Driven Learning for Programs} that uses abstract model search and abstract
conflict analysis procedures for safety verification of C programs.  The
model search procedure operates on an over-approximate domain of program
traces through repeated application of abstract deduction transformer,
$\abstransel{}$, and decisions in order to search for a counterexample
trace.  If the model search finds a satisfying assignment (corresponding
deduction transformer is $\gamma$-complete), then ACDLP terminates with a
counterexample trace, and the program is \emph{unsafe}.  Else, if a conflict
is encountered, then it implies that the corresponding program trace is
either not valid or safe.  ACDLP then moves to the conflict analysis phase
where it learns the reason for the conflict from partial safety proof using
an abstract abductive transformer, $\abd{}{}$, followed by a heuristic
choice of conflict reason.  Similar to a SAT solver, ACDLP picks one
conflict reason from multiple incomparable reasons for conflict for
efficiency reasons.  Hence, it operates over an under-approximate domain of
conflict reasons.  A~conflict reason under-approximates a set of invalid or
safe traces.  The conflict analysis returns a learnt transformer (negation
of conflict reason) that over-approximates a set of valid and unsafe traces. 
Model search is repeated with this new transformer.  Else, if no further
backtracking is possible, then ACDLP terminates and returns \emph{safe}. 
We~present the ACDLP algorithm in subsequent section.

The input to ACDLP (Algorithm~\ref{Alg:acdcl}) is a
program in the form of a set of abstract transformers
$\abstransset=\{\abstrans{\domain}{\sigma}|\sigma\in\Sigma\}$
w.r.t.\ an abstract domain~$\domain$.  Recall that the safety 
formula $\bigwedge_{\constraint\in\constraints} \constraint$ 
is unsatisfiable if and only if the program is safe.  
The algorithm is parametrised by heuristics for propagation $(\propheur)$, 
decisions $(\decheur)$, and conflict analysis $(\confheur)$.
The algorithm maintains a propagation trail $\trail$ and 
a reason trail~$\reasons$.
The propagation trail stores all meet irreducibles inferred by 
the abstract model search phase (deductions and decisions).  
The reason trail maps the elements of the propagation trail to the
transformers $\abstransel{}\in\abstransset$ that were used to
derive them. 
\begin{definition} 
The \emph{abstract value} $\abs(\trail)$ corresponding to 
the propagation trail $\trail$ is the conjunction of the 
meet irreducibles on the trail:
$\abs(\trail)=\bigsqcap_{m \in \trail}m$ with
$\abs(\trail)=\top$ if $\trail$ is the empty sequence.
\end{definition}
The algorithm begins with an empty $\trail$, an empty $\reasons$, and the
abstract value $\top$.  The procedure $\deduce$ (details in
Section~\ref{sec:deduce}) computes a greatest fixed-point over the
transformers in $\abstransset$ that refines the abstract value,
similar to the Boolean Constraint Propagation
step in SAT solvers.  If the result of $\deduce$
is \textsf{conflict} ($\bot$), the algorithm terminates with
\textsf{safe}.  Otherwise, the analysis enters into the while loop at line 4
and makes a new decision by a call to $\decide$ (see
Section~\ref{sec:decide}), which returns a new meet irreducible
$\decisionvar$.
%
%
We concatenate $\decisionvar$ to the trail~$\trail$.  The decision
$\decisionvar$ refines the current abstract value $\abs(\trail)$ represented
by the trail, i.e., $\abs(\trail\cdot\decisionvar)\sqsubseteq \abs(\trail)$.
%
%
For example, a decision in the interval domain restricts the range of 
intervals for variables.
We set the corresponding entry in the reason trail~$\reasons$ to $\top$
to mark it as a decision.  Here, the index of $\reasons$ is the size 
of trail $\trail$, denoted by $|\trail|$.
The procedure $\deduce$ is called next to infer new meet irreducibles
based on the current decision.  The model search phase
alternates between the decision and deduction until $\deduce$ returns
either \textsf{sat} or \textsf{conflict}.  

If  $\deduce$ returns  \textsf{sat}, then we have found an abstract value 
that represents models of the safety formula, which are counterexamples 
to the required safety property, and so ACDLP returns \textsf{unsafe}.
%
%
If  $\deduce$ returns  \textsf{conflict}, 
the algorithm enters in the $\analyzeconflict$ 
phase (see Section~\ref{sec:conflict}) to learn the reason for the conflict.   There can be multiple
incomparable reasons for conflict.
ACDLP heuristically chooses one reason~$\conflictset$ and learns it 
by adding it as an abstract transformer to $\abstransset$. The analysis 
backtracks by removing the content of $\trail$ up to a point where it does not 
conflict with $\conflictset$.  ACDLP then performs deductions with the learnt 
transformer.  If $\analyzeconflict$ returns $\false$, then no further
backtracking is possible.  Thus, the safety formula is unsatisfiable
and ACDLP returns \textsf{safe}.  
\section{Abstract Model Search for Template Polyhedra}\label{sec:deduce}

\Omit{
\begin{figure}[t]
\scriptsize
\begin{tabular}{l|l|l}
\hline
C program & SSA & Octagon \\
\hline
\begin{lstlisting}[mathescape=true,language=C]
int main() {
 unsigned x, y;
 __CPROVER_assume(x==y);
 x++;
 assert(x==y+1);
}
\end{lstlisting}
&
\begin{minipage}{4.40cm}
$\begin{array}{l@{\,\,}c@{\,\,}l}
SSA &\iff& ((g0 == TRUE) \land \\
    &    & (cond == (x == y)) \land \\
    &    & (g1 == (cond \&\& guard0)) \land \\
    &    & (x' == 1u + x) \land \\
    &    & (x' == 1u + y || !1))
\end{array}$
\end{minipage}
&
\begin{minipage}{3.75cm}
$\begin{array}{l@{\,\,}c@{\,\,}l}
C &\iff& ((x' > 1) \land (-x'-y < -2) \land \\
  &    & (-x-x' < -2) \land (y-x' < 0) \land \\                                                                
  &    & (x-x' < 0) \land (y > 0) \land \\
  &    & (x > 0) \land (-x'-y < 0) \land \\
  &    & (x+y > 1) \land (y-x < 1) \land \\
  &    & (x'-y < 2) \land (x-y < 1) \land \\
  &    & (x+y > 0) \land (x+x' > 0) \land \\
  &    & (x'-x < 2))
\end{array}$
\end{minipage}
\\
\hline
\end{tabular}
\caption{C Program, corresponding SSA and Octagonal Inequalities}
\label{ssa}
\end{figure}
}

\begin{algorithm2e}[t]
\DontPrintSemicolon
\SetKw{return}{return}
\SetKwData{sat}{sat}
\SetKwData{conflict}{conflict}
\SetKwData{unsat}{unsat}
\SetKwData{unknown}{unknown}
\SetKwData{true}{true}
\SetKwInOut{Input}{input}
\SetKwInOut{Output}{output}
\SetKwFor{Loop}{Loop}{}{}
\SetKw{KwNot}{not}
\begin{small}
\Input{A program in the form of a set of abstract transformers $\abstransset$,
a propagation trail $\trail$, and a reason trail $\reasons$.}
\Output{\sat or \conflict or \unknown}
$\worklist \leftarrow \initworklist_{\propheur}(\abstransset)$ \;
\While{$!\mathit{worklist.empty}()$} 
{
  $\abstransel{\subdomain} \leftarrow \mathit{worklist.pop}()$ \; 
  $\absval \leftarrow \abstransel{\subdomain}(\abs(\trail))$\;
  \uIf{$\absval = \bot$} {
    $\reasons[\bot] \leftarrow \abstransel{\subdomain}$ \;
    $\mathit{worklist.clear}()$ \;
    \return \conflict \;
  }
  \uElse
  {
    $\newdeductions=\onlynew(\absval)$\;
    $\trail \leftarrow \trail \cdot \decomp(\newdeductions)$ \; 
    $\reasons[|\trail|] \leftarrow \abstransel{\subdomain}$ \;
    $\updateworklist_{\propheur}(\worklist, \newdeductions, \abstransel{\subdomain},  \abstransset)$ \; 
  }
}
\lIf{$\abstransset$ is $\gamma$-complete at $\abs(\trail)$} {
  \return \sat
}
 \return \unknown \;

\end{small}
\caption{Abstract Model Search $\mathit{deduce}_{\propheur}(\abstransset,\trail,\reasons)$ \label{Alg:ms}}
\end{algorithm2e}

Model search in a SAT solver has two steps: {\em deductions}, which are
repeated application of the unit rule (also called Boolean Constraint
Propagation, or BCP), to refine current partial assignments, and {\em
decisions} to heuristically guess a value for an unassigned literal.  BCP
can be seen to compute greatest fixed point over the partial assignment
domain~\cite{dhk2013-popl}.  Below, we present an abstract model search
procedure that computes a greatest fixed point over abstract
transformers~$\abstrans{\domain}{\constraint}$.
 
%

\subsection{Parametrised Abstract Transformers} \label{sec:abst}
\Omit{
To make our algorithm efficient, we have to focus the abstract
transformers on performing only the minimally necessary
work. 
}

The key considerations for an abstract transformer are precision and
efficiency.  A~precise transformer is usually less efficient than a more
imprecise one.  In this paper, we present a specialised variant of the
abstract transformer to compute deductions called \emph{Abstract Deduction
Transformer} (ADT), which is parametrised by a given \emph{subdomain}
$\subdomain\subseteq \domain$.
A subdomain contains a chosen subset of the elements in $\domain$ including
$\bot$ and $\top$ that forms a lattice.
The use of a subdomain serves two purposes -- 
  a) It allows us elegantly and flexibly to guide the deductions in 
  {\em forward}, {\em backward} or {\em multi-way} direction, which 
  in turn affects the analysis precision, and 
  b) It makes deductions more efficient, for example by performing lazy closure
  in template polyhedra domain. 
  For space reasons, we refer the reader to Appendix~\ref{appendix:lazyclosure} 
  for details of the lazy closure operation. 
%

An ADT is defined formally  as follows. 
\begin{equation}\label{eq:at2}
\abstrans{\domain}{\constraint}^\subdomain(\absval)=\absval\meet_\domain \alpha_\subdomain(\{\val\mid \val\in\gamma_\domain(\absval), \val\models \constraint\})
\end{equation}
For $\subdomain=\domain$, the ADT is
identical to the abstract transformer defined in
Eq.~(\ref{eq:abstrans}) in Section~\ref{sec:domains}.  Note 
that a restricted subdomain makes a transformer less 
precise but more efficient.  Conversely, an
unrestricted subdomain make a transformer more precise, but less
efficient. Therefore, we have the property
$\abstrans{\domain}{\constraint}^\domain(\absval)\sqsubseteq
\abstrans{\domain}{\constraint}^\subdomain(\absval)$.
\Omit{Thus, the
parametrisation helps us fine-tune the precision and efficiency of
deductions.}
\Omit{Furthermore, the choice of subdomain internally
  guides the deduction in {\em forward}, {\em backward} or {\em
    multi-way} direction which are described next.}
To illustrate point (1), 
we give examples that demonstrate how the choice of
subdomain influences the propagation direction:

\paragraph{Forward Transformer.} 
For an abstract value
$\absval=(0\leq y \leq 1 \wedge 5\leq z)$, $\constraint=(x=y+z)$, 
and $L={\intervals[\{x\}]}$,  we have
$\abstrans{\intervals[\{x,y,z\}]}{x=y+z}^{\intervals[\{x\}]}(\absval)=\absval\meet(x\geq
6)$.
Assuming that the equality $x=y+z$ originated from an assignment to $x$,
this performs a right-hand side (rhs) to left-hand side (lhs) propagation and
hence emulates a forward analysis. 

\paragraph{Backward Transformer.} 
For an abstract value $\absval=(0\leq x \leq 10 \wedge 0\leq y \leq 1 \wedge 5\leq z)$, 
$\constraint=(x=y+z)$, and $L={\intervals[\{y,z\}]}$, we have
$\abstrans{\intervals[\{x,y,z\}]}{x=y+z}^{\intervals[\{y,z\}]}=\absval\meet (z\leq 10)$. 
This performs an lhs-to-rhs propagation and hence emulates a backward analysis.  

\paragraph{Multi-way Transformer.} 
\Omit{We call the propagation with arbitrary subdomains,e.g. 
$\subdomain=\domain$, \emph{multi-way propagation}, which is able
to simultaneously perform forward and backward propagation.}
For an abstract value $\absval=(c\leq1 \wedge c\geq1 \wedge x\leq5 \wedge x\geq5)$, 
$\constraint=((c = (x=y)) \wedge y=y+1)$ and $L={\intervals[\{c,x,y\}]}$, we have 
$\abstrans{\intervals[\{c,x,y\}]}{\constraint}^{\intervals[\{c,x,y\}]}=\absval\meet(y\leq6
\wedge y\geq6)$.  This performs an lhs-to-rhs propagation for $c=(x=y)$ and rhs to lhs propagation
for $y=y+1$ and hence emulates a multi-way analysis.  
\subsection{Algorithm for the Deduction Phase}
%
Algorithm~\ref{Alg:ms} presents the deduction phase $\deduce$ in 
our abstract model search procedure.  The input to $\deduce$ is 
the set of abstract transformers, a propagation trail ($\trail$) 
and a reason trail~($\reasons$).  Additionally, the procedure 
$\deduce$ is parametrised by a propagation heuristic ($\propheur$). 
We write the ADT 
$\abstrans{\domain}{\constraint}^\subdomain$
as $\abstransel{\subdomain}$ in Algorithm~\ref{Alg:ms}. 
The algorithm maintains a {\em worklist}, which is a queue that contains 
ADTs.  The propagation heuristics provides two 
functions $\initworklist$ and $\updateworklist$.
The order of the elements in the worklist and the subdomain $\subdomain$ 
associated with each ADT ($\abstransel{\subdomain}$) 
determine the propagation strategy (forward, backward, multi-way).
These two functions construct a subdomain ($\subdomain$)
for $\abstransel{\subdomain}$ 
by calling the function $\makesubdomain$ such that 
$\subdomain=\makesubdomain_\domain(\subvars)$, where $\subvars$ are 
the variables that appear in $\abstransel{\subdomain}$. 
\Omit{
The {\em forward} propagation strategy initialises the worklist with
ADTs that contain constants in the
right-hand side; the subdomain is constructed via $\makesubdomain_\domain$
from variables in the left-hand side of the equality constraints
originating from the assignment statements in the program.
The {\em backward} propagation strategy initialises the worklist 
with the assertions; the subdomain is constructed from the 
right-hand side variables.
The {\em multi-way} propagation strategy initialises the worklist 
with the set of all transformers; corresponding subdomains 
contains the variables occurring in the respective transformers.
}
The abstract value $\absval$ is updated upon the application of 
$\abstransel{\subdomain}$ in line~4 in Algorithm~\ref{Alg:ms}. 
The function
$\onlynew(\absval)=\bigsqcap(\decomp(\absval)\setminus\decomp(\abs(\trail)))$
is used to filter out all meet irreducibles that are already on the trail
in order to obtain only new deductions ($\newdeductions$) when applying 
the ADT (shown in line~10).
Depending on the propagation heuristics, $\updateworklist$ adds
ADTs $\abstransel{\subdomain}$ to the 
worklist that contain variables that appear in $\newdeductions$, and 
updates the subdomains of the ADTs in the worklist 
to include the variables in $\newdeductions$ (shown in line~13).
\Omit{
If $\bot$ is deduced we return \textsf{conflict}.
Otherwise, when eventually a fixed point has been reached, i.e.\ the worklist is empty, then the abstract value $\abs(\trail)$ is checked whether it is 
$\gamma$-complete~\cite{dhk2013-popl}. 
It is $\gamma$-complete if all concrete values in $\gamma(\abs(\trail))$ satisfy $\formula$.
Otherwise, the algorithm returns \textsf{unknown} and ACDLP makes a new decision.}

If $\abstransel{\subdomain}$ deduces $\bot$, then 
the procedure $\mathit{deduce}$ returns \textsf{conflict} (shown in line~8).
Otherwise, when a fixed-point is reached, i.e.~the worklist is empty, we check whether
the abstract transformers $\abstransset$ are $\gamma$-complete~\cite{dhk2013-popl} for the current abstract value $\abs(\trail)$ 
(shown in line~15).
Intuitively, this checks whether all concrete values in 
$\gamma(\abs(\trail))$ satisfy the safety formula $\formula$, where 
$\formula:= \bigwedge_{\constraint\in\constraints} \constraint$ is obtained 
from the program transformation (as defined in Section~\ref{sec:program}).
If it is indeed 
$\gamma$-complete, then $\mathit{deduce}$ returns \textsf{sat}.  Otherwise, the 
algorithm returns \textsf{unknown} and ACDLP makes a new decision.    
\subsection{Computing Lazy Closure for Template Polyhedra}\label{lazyclosure}
%
An advantage of our formalism in Eq.~(\ref{eq:at2}) is that the 
\emph{closure} operation for relational domains can be computed 
in a lazy manner through the construction of a subdomain, $\subdomain$.  
The construction of $\subdomain$ allows us to perform one step of the 
closure operation when $\abstransel{\subdomain}$ is applied.
For example, let us consider $\domain=\octagons[\{x,y,z\}]$ and
$\subvars=\{y\}$. An octagonal inequality relates at 
most two variables. Thus it is sufficient to consider the subdomain
$\makesubdomain_\domain(\{y\})=\octagons[\{y\}]\cup\octagons[\{x,y\}]\cup\octagons[\{y,z\}]$,
which will compute the one-step transitive relations of~$y$ with each
of the other variables. 
Only if any subsequent abstract deduction transformer makes new deductions 
on $x$ or $z$, then the next step of the closure will be computed through 
the subdomain $\octagons[\{x,z\}]$.
\Omit{
Hence, when the abstract deduction transformer is applied we do not
compute the full closure in the full domain,
but we compute only a single step of the closure in a restricted
domain, which makes single deduction steps more efficient.
}
Hence, an application of each abstract deduction transformer does not 
compute the full closure in the full domain, but compute only a single 
step of the closure in a subdomain. This makes each deduction step more 
efficient but may require more steps to reach the fixed point.  

%


\subsection{Decisions}\label{sec:decide}
A decision $\decisionvar$ is a meet irreducible that refines the
current abstract value $\abs(\trail)$, when the result of fixed-point 
computation through deduction is neither a {\em conflict} nor a {\em 
satisfiable model} 
of $\formula$.  A decision must always be consistent 
with respect to the trail $\trail$, 
i.e., $\abs(\trail\cdot \decisionvar)\neq \bot$.  A new 
decision increases the decision level by one. Given the 
current abstract value $\abs(\trail)$, the procedure $\decide$ 
in Algorithm~\ref{Alg:acdcl} heuristically returns a meet irreducible.

%
For example, a decision in the interval domain can be of the form 
$x R d$ where $R \in \{\leq,\allowbreak\geq\}$, and $d$ 
is the bound.  A decision in the octagon domain can specify relations 
between variables, and can be of the form $ax - by \leq d$, where 
$x$ and $y$ are variables, $a,b \in \{-1,0,1\}$ are coefficients, 
and $d$ is a constant.  We refer the reader to Appendix~\ref{decision} 
for the details of different decision heuristics in ACDLP.

\section{Abstract Conflict Analysis for Template Polyhedra}\label{sec:conflict}
Propositional conflict analysis with FIRST-UIP~\cite{cdcl} can be seen as
abductive reasoning that under-approximates a set of models that do not
satisfy a formula~\cite{sas12,dhk2013-popl}.  Below, we present an abstract
conflict analysis procedure, $\analyzeconflict$ of
Algorithm~\ref{Alg:acdcl}, that uses a domain-specific abductive transformer
for effective learning.  A~conflict analysis procedure involves two steps:
{\em abduction} and {\em heuristic choice for generalisation}.  Abduction
infers possible generalised reasons for a conflict which is followed by
heuristically selecting a generalisation.  Below, we define a global 
conflict transformer that gives a set of models that do not satisfy a 
formula.  
\begin{definition}
  Given formula $\formula$, an initial abstract value $a$ and domain 
  $\domain$, $\conf{\formula}{\domain}(a) = \{u \mid u \in decomp(a) 
  \vee u \not\models \formula \}$, that is, it adds abstract models to 
  the input set $a$, that do not satisfy $\formula$.
\end{definition}
For formula $\formula$ and domain $\domain$, an abstract abductive transformer, 
$\abd{\formula}{\domain}(a)$, corresponds 
to the under-approximation of the global conflict transformer,
$\conf{\formula}{\domain}$.  
For example, given a formula $\formula=\{x=y+1\; \&\&\; x\geq 0\}$, and an interval 
abstract element $a=(y\leq-5)$, $\conf{\formula}{\intervals}(a) = \{(y\leq-5), 
(y \leq -4, x \leq -3), (y \leq -3, x \leq -2), (y \leq -2, x < 0)\}$.
Informally, $\conf{\formula}{\intervals}$  computes the most general set of
incomparable reasons under which $\formula$ implies the truth of $a$ (or $\bot$
since $formula$ is unsatisfiable under $a$). 
Now, an abstract abductive transformer for $\formula=\{x=y+1\; \&\&\; x \geq 0\}$ 
is given by, $\abd{\formula}{\intervals}(y \leq -5)=(y \leq -2, x < 0)$,
which clearly underapproximates $\conf{\formula}{Itvs}$ as well as strictly
generalizes the reason for $y \leq -5$. 

The main idea of abductive  reasoning is to iteratively replace a singleton 
assignment $s$ in the conflict reason by a partial assignment that is sufficient 
to infer $s$.  Conflict abduction is performed by obtaining cuts through markings 
in the trail $\trail$, by the application of abstract Unique Implication Point (UIP) 
search algorithm~\cite{cdcl}.  Every cut is a reason for conflict.  The UIP
search can also be understood through graph cutting in an Abstract Conflict
Graph, which is defined next. 
\begin{definition}
  An Abstract Conflict Graph (ACG) is a directed acyclic graph in which the 
  vertices are defined by all deduced elements or a decision node and a special 
  conflict node $(\bot)$ in the trail $\trail$.  The edges in ACG are obtained 
  from the reason trail $\reasons$ that maps the elements in $\trail$ to the abstract
  transformers that are used to derive the deduced elements.  
  \Omit{
  if $\abstrans{\domain}{\constraint}(a) = 
  u$, where $u$ is an unique deduction from $\abstrans{\domain}{\constraint}$
  and $a$ is an initial abstract value, then there is a directed edge from each 
  elements of $a$, whose variable set has non-empty intersection with the variable 
  set of $\constraint$, to $u$. 
  }
\end{definition}
\begin{figure}[t]
\centering
\scalebox{.6}{\import{figures/}{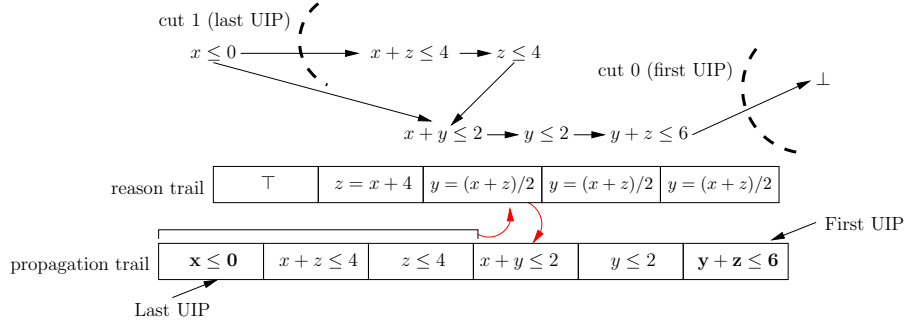}}
\caption{\label{uip}Finding the Abstract UIP in the Octagon Domain}
\end{figure}  
\noindent \textbf{Abstract UIP Search}
An abstract UIP algorithm~\cite{DBLP:journals/fmsd/BrainDGHK14} traverses
the trail $\trail$ starting from the conflict node and computes a cut that
suffices to produce a conflict.  For example, consider a formula $\formula:=
\{x{+}4{=}z \wedge x{+}z{=}2y \wedge z{+}y>10\}$.  As before, the trail can be
viewed to represent an ACG, shown in Fig.~\ref{uip}, that records the 
sequence of deductions in the octagon domain that are inferred
from a decision $(x{\leq}0)$ for the formula $\formula$.  The arrows (in
red) shows the relationship between the reason trail and propagation trail
in bottom of Fig.~\ref{uip}.  For the partial abstract value,
$\absval=\{x\leq0 \wedge x+z\leq4 \wedge z\leq4\}$, obtained from the trail,
the result of the abstract deduction transformer is
$\abstrans{\octagons}{y=(x+z)/2}(\absval)=\{x+y\leq2, y\leq2, y+z\leq6\}$. 
A~conflict ($\bot$) is reached for the decision ${x{\leq}0}$.  Note that
there exist multiple incomparable reasons for conflict, marked as {\em cut0}
and {\em cut1} in Fig.~\ref{uip}.  Here, cut0 is the first UIP (node closest
to conflict node).  Choosing cut0 yields a learnt clause $(y+z>6)$, which is
obtained by negating the reason for conflict.  The abstract UIP algorithm
returns a learnt transformer $\aunit$, which is described next.\\

\noindent \textbf{Learning in Template Polyhedra Domain}
Learning in a propositional solvers yields an asserting clause~\cite{cdcl}
that expresses the negation of the conflict reasons.  We present a
lattice-theoretic generalisation of the {\em unit rule} for template-based
abstract domains that learns a new transformer called {\em abstract unit
transformer} $(\aunit)$.  We add $\aunit$ to the set of abstract
transformers $\abstransset$.  $\aunit$ is a generalisation of the
propositional unit rule to numerical domains.  For an abstract lattice
$\domain$ with complementable meet irreducibles and a set of meet
irreducibles $\conflictset \subseteq \domain$ such that $\bigsqcap
\conflictset$ does not satisfy $\formula$, $\aunit_\conflictset: \domain
\rightarrow \domain$ is formally defined as follows.
\[ \aunit_\conflictset(\absval) =
 \left\{\begin{array}{l@{\quad}l@{\qquad}l}
  \bot       & \text{if } \absval \sqsubseteq \bigsqcap \conflictset & (1)\\
  \bar{t}    & \text{if } t \in \conflictset \; \text{and} \; \forall t' \in \conflictset
  \setminus \{t\}. \absval  \sqsubseteq t' & (2) \\
  \top & \text{otherwise} & (3) \\
 \end{array}\right.
\]
Rule (1) shows $\aunit$ returns $\bot$ since $\absval \sqsubseteq \bigsqcap
\conflictset$ is conflicting.  Rule (2) of $\aunit$ infer a valid meet
irreducible, which implies that $\conflictset$ is unit.  Rule (3) of
$\aunit$ returns $\top$ which implies that the learnt clause is not {\em
asserting} after backtracking.  This would prevent any new deductions from
the learnt clause.  Progress is then made by decisions.  An example of
$\aunit$ for $\conflictset = \{x \geq 2, x \leq 5, y \leq 7 \}$ is given
below.  \\
\Omit{
Let us consider an example, where $\conflictset = \{x \geq 2, x \leq 5, y
\leq 7 \}$ and $\absval = (x \geq 3\wedge\allowbreak x \leq
4\wedge\allowbreak y \geq 5\wedge y \leq 6)$.  Then
$\aunit_\conflictset(\absval) = \bot$ using the rule (1), since $\absval
\sqsubseteq \bigsqcap\conflictset$.  Now, consider another abstract value
$\absval = (x \geq 3\wedge x \leq 4)$ and the same $\conflictset$ as above,
then $\aunit_\conflictset(\absval) = (y \geq 8)$ using rule (2), since
$\absval \sqsubseteq (x \geq 2)$ and $\absval \sqsubseteq (x \leq 5)$. \\ 
}
Rule 1: 
For $\absval = (x \geq 3\wedge\allowbreak x \leq
4\wedge\allowbreak y \geq 5\wedge y \leq 6)$, 
$\aunit_\conflictset(\absval) = \bot$, since $\absval
\sqsubseteq \bigsqcap\conflictset$.  \\ 
Rule 2:  
For $\absval = (x \geq 3\wedge x \leq 4)$,  
$\aunit_\conflictset(\absval) = (y \geq 8)$, since
$\absval \sqsubseteq (2 \leq x \leq 5)$. \\ 
Rule 3: 
For $\absval = (x \geq 1\wedge y \leq 10)$, 
$\aunit_\conflictset(\absval) = \top$. \\

\noindent \textbf{Backjumping}
A backjumping procedure removes all the meet irreducibles from 
the trail up to a decision level that restores the analysis to a
non-conflicting state.  The backjumping level is defined by the
meet irreducibles of the conflict clause that is closest 
to the root (decision level~0) where the conflict
clause is still unit.  If a conflict clause is globally unit, then the
backjumping level is the root of the search tree and
$\analyzeconflict$ returns $\false$, otherwise it returns $\true$.


\section{Experimental Results}
We have implemented ACDLP for bounded safety verification of C programs.  
ACDLP is implemented in C++ on top of the
\textsc{CPROVER}~\cite{cprover} framework and consists of around 9~KLOC. 
The template polyhedra domain is implemented in C++ in 10~KLOC.  Templates
can be intervals, octagons, zones, equalities, or restricted polyhedra.  Our
domain handles all C operators, including bit-wise ones, and supports
precise complementation of meet irreducibles, which is necessary for
conflict-driven learning.  Our tool and benchmarks are available 
at~\url{http://www.cprover.org/acdcl/}.

We verified a total of~85 ANSI-C benchmarks.  These are derived from:
(1)~the bit-vector regression category in SV-COMP'16; (2)~ANSI-C models of
hardware circuits auto-generated by v2c~\cite{mtk2016} from VIS Verilog
models and opencores.org; (3)~controller code with varying loop bounds 
auto-generated from Simulink model and control 
intensive programs with nested loops containing relational properties. 
All the programs with bounded loops are completely unrolled before
analysis.

We~compare ACDLP with the state-of-the-art SAT-based bounded model checker
CBMC (\cite{cbmc}, version 5.5) and a commercial static analysis tool,
Astr{\'e}e (\cite{astree}, version 14.10).  CBMC uses MiniSAT~2.2.1 in the
backend.  Astr{\'e}e uses a range of abstract domains, which includes
interval, bit-field, congruence, trace partitioning, and relational domains
(octagons, polyhedra, zones, equalities, filter).  To enable fair comparison
using Astr{\'e}e, all bounded loops in the program are completely unwound up
to a given bound before passing to Astr{\'e}e.  This prevents Astr{\'e}e
from widening loops.
ACDLP is instantiated to a product of the Booleans and the interval or
octagon domain instance of template polyhedra.  ACDLP is also
configured with a decision heuristic (ordered, random,
activity-based), propagation (forward, backward and multi-way), and
conflict-analysis (learning UIP, DPLL-style).  The timeout for our
experiments is set to~200 seconds.
\Omit {
To enable precise analysis using Astr{\'e}e, all our benchmarks are 
manually instrumented with partition directives which provides external 
hint to the tool to guide the trace partitioning heuristics.  Usually, 
such high-precision is not needed for static analysis, since it makes 
the analysis very expensive.  Without trace partitioning, the 
analysis using Astr{\'e}e shows high degree of imprecision. 
}

\begin{figure}[t]
  \centering
\begin{tabular}{@{\hspace{-1.5em}}c@{\hspace{1em}}c}
\begin{tikzpicture}[scale=0.60]
	\begin{loglogaxis} [xmin=0.1,xmax=4000, ymin=0.1, ymax=4000, xlabel= SAT
    (Decisions),
			ylabel=ACDLP (Decisions), 
			legend pos = north west,
			]
\addplot [only marks,scatter,point meta=explicit symbolic,
	scatter/classes={s={mark=square,mark size=2.5},u={mark=triangle*,blue,mark size=2.5}},] 
	table [meta=label] {plotdata/scatter-decision.dat};
	\legend{Safe,Unsafe}
\addplot [domain=.1:4000] {x};
\end{loglogaxis}
\end{tikzpicture}
 &
\begin{tikzpicture}[scale=0.60]
  \centering
	\begin{loglogaxis} [xmin=.1,xmax=83000, ymin=.1, ymax=83000, xlabel=SAT
    (Propagations),
			ylabel=ACDLP (Propagations),
			legend pos = north west,
			]
\addplot [only marks,scatter,point meta=explicit symbolic,
	scatter/classes={s={mark=square,mark size=2.5},u={mark=triangle*,blue,mark size=2.5}},] 
	table [meta=label] {plotdata/scatter-propagation.dat};
	\legend{Safe,Unsafe}
\addplot [domain=.1:83000] {x};
\end{loglogaxis}
\end{tikzpicture} \\
(a) & (b)
\end{tabular}
\caption{\label{fig:results}
Comparison between SAT-based BMC and ACDLP: number of decisions and propagations}
\end{figure}
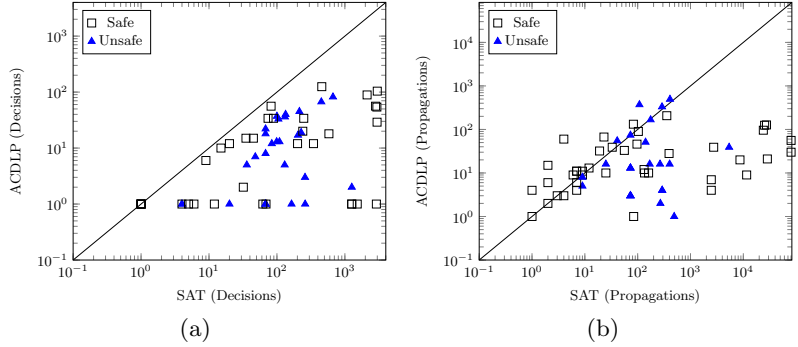

\begin{figure}[t]
  \centering
  \begin{tikzpicture}[scale=0.60]

\pgfplotscreateplotcyclelist{markstyles}{%
solid, every mark/.append style={solid, fill=white}, mark=square*, mark size=2.5\\%
solid, every mark/.append style={solid, blue}, mark=triangle*,mark size=2.5\\%
solid, every mark/.append style={solid, fill=black}, mark=otimes*,, mark
    size=2.5\\%
}
    
  \begin{axis}[
    width=\linewidth,
    xlabel={Benchmark Number},
    ylabel={Time (seconds)},
    domain = 1:85,
    xmin=1, xmax=85,
    ymin=0, ymax=200,
    xtick={1,5,10,...,85},
    width=20cm, height= 7cm,
    ymode = log,
    legend pos = north west,
    grid = major,
    major grid style={line width=.2pt,draw=gray!50},
    cycle list name=markstyles
  ]
	
    \addplot table [only marks, y=Time, x=Benchmarks]{plotdata/cbmc.dat};
	\addlegendentry{CBMC}
  
  \addplot table [only marks, y=Time, x=Benchmarks]{plotdata/acdlp.dat};
  \addlegendentry{ACDLP}
  
  \addplot table [only marks, y=Time, x=Benchmarks]{plotdata/astree.dat};
    \addlegendentry{Astr{\'e}e}
	
  \end{axis}  
\end{tikzpicture}
\caption{\label{fig:runtimes}
  Runtime Comparison between CBMC, Astr{\'e}e and ACDLP}
\end{figure}
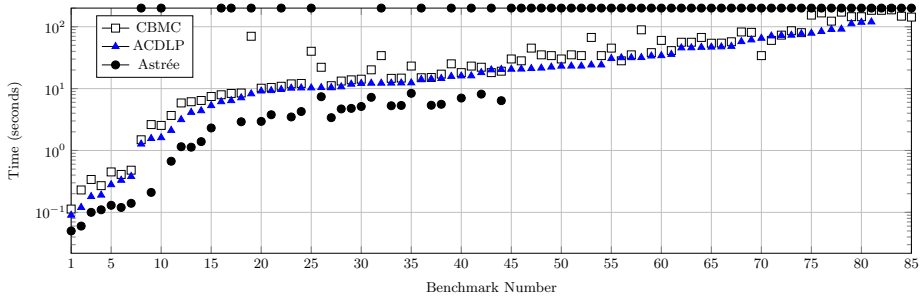

\noindent \textbf{ACDLP versus CBMC}
Fig.~\ref{fig:results} presents a comparison of the analyses using CBMC
and ACDLP.  Fig.~\ref{fig:results}(a) clearly shows that the SAT based analysis 
made significantly more decisions compared to ACDLP for all the benchmarks. 
The points on the extreme right below the diagonal in
Fig.~\ref{fig:results}(b) show that the number of propagations in the SAT based 
analysis is maximal for benchmarks that exhibit relational behaviour.  These
benchmarks are solved by octagon domain in ACDLP.  We see a reduction of at 
least two orders of magnitude in the total number of decisions, propagations 
and conflicts compared to analysis using CBMC.  

Out of 85 benchmarks, SAT based analysis could prove only 26
benchmarks without any restarts.  The solver was restarted in the other 59 
cases to avoid spending too much time in ``hopeless'' branches.  By contrast, 
ACDLP solved all 85 benchmarks without restarts.  
The runtime comparison between ACDLP and CBMC are shown in 
Figure~\ref{fig:runtimes}.  ACDLP is~1.5X faster than CBMC. 
The superior performance of ACDLP is attributed to the decision heuristics, 
which exploit the high-level structure of the program, combined with the 
precise deduction by multi-way transformer and stronger learnt clause aided 
by the richer abstract domains. 

\noindent \textbf{ACDLP versus Astr{\'e}e}
To enable precise analysis using Astr{\'e}e, we manually instrument 
the benchmarks with partition directives \texttt{\_\_ASTREE\_partition\_control} 
at various control-flow joins.  These directives provide external hint to
Astr{\'e}e to guide its internal trace partition domain. 
Figure~\ref{fig:runtimes} demonstrates that Astr{\'e}e is~2X faster 
than ACDLP for {37}\% cases (32 out of 85); but the analysis using 
Astr{\'e}e shows a high degree of imprecision (marked as timeout in 
Figure~\ref{fig:runtimes}).  Astr{\'e}e reported~53 false alarms 
among~85 benchmarks.  Whereas, the analysis using ACDLP produces correct 
results for~81 benchmarks.  ACDLP timed out for~4 benchmarks. Clearly, 
ACDLP has higher precision than Astr{\'e}e. Detailed analysis of the 
comparison between ACDLP, CBMC and Astr{\'e}e is presented in 
Appendix~\ref{appendix:extended_result}.

Our experimental evaluation suggests that ACDLP can be seen as a
technique to improve the efficiency of SAT-based BMC.  Additionally, ACDLP can
also be perceived as an automatic way to improve the precision of conventional
abstract interpretation over non-distributive lattices through automatic partition 
generation techniques such as decisions and transformer learning.


\section{Related Work}
The work of~\cite{franzle} presents a tight integration of SAT solving
with interval based arithmetic constraint solving to handle large constraint
systems. 
Silva et al.~\cite{sas12} present an abstract interpretation account of 
satisfiability algorithms derived from DPLL procedures.  
The work of~\cite{tacas12} is a very early instantiation of abstract 
CDCL~\cite{sas12} as an interval-based decision procedure for programs, 
but in a purely logical settings.  
A similar technique that lifts DPLL(T) to programs is Satisfiability Modulo 
Path Programs (SMPP)~\cite{SMPP}. SMPP enumerates program paths using a SAT 
formula, which are then verified using abstract interpretation.  
The work of \cite{DBLP:conf/esop/MineBR16} proposes an algorithm inspired by 
constraint solvers for inferring disjunctive invariants using intervals.
The lifting of CDCL to first-order theories is proposed
in~\cite{dpll,ndsmt,vmcai13}.
\Omit{ operates on a fixed first-order partial assignment lattice
  structure, where first-order variables are mapped to domain values,
  similar to constants lattice in program analysis.  } 
  Unlike previous work that operates on a fixed first-order lattice,
  ACDLP can be instantiated with different abstract domains as well as {product
  domains}.  
  \Omit{This involves model search and learning in abstract lattices.  A similar
  technique that lifts decisions, propagations and learning to theory variables
  is Model-Constructing Satisfiability Calculus (mcSAT)~\cite{vmcai13}.}
  
  ACDLP is not, however, similar to abstraction refinement. ACDLP works on a fixed
  abstraction. Also, transformer learning in ACDLP does not soundly over-approximate
  the existing program transformers. Hence, transformer learning in ACDLP is
  distinct from transformer refinement in classical CEGAR. 

\Omit { The
  abstract lattice in natural domain SMT does not have complementable
  meet irreducibles, and therefore does not support generalized clause
  learning~\cite{sas}.  On the other hand, most abstract lattice have
  complementation property, thus enabling ACDCL to perform generalized
  clause learning.  } 
%
%
%
%


\section{Conclusions}

In this paper, we present a general algorithmic framework for lifting the
model search and conflict analysis procedures in satisfiability solvers to
program analysis.  We embody these techniques in a tool, ACDLP, for
automatic bounded safety verification of C programs over a template
polyhedra abstract domains.

We present an {\em abstract model search} procedure that uses a
parameterised abstract transformer to flexibly control the precision and
efficiency of the deductions in the template polyhedra abstract domain.  The
underlying expressivity of the abstract domain helps our decision heuristics
to exploit the high-level structure of the program for making effective
decisions.  The {\em abstract conflict analysis} procedure learns abstract
transformers over a given template following a UIP computation. 
Experimental evaluation over a range of benchmarks shows~20x reduction in
the total number of {\em decisions}, {\em propagations}, {\em conflicts} and
{\em backtracking} iterations compared to CBMC.  Moreover, ACDLP is~1.5x
faster than CBMC.  Compared to Astr{\'e}e, ACDLP solves twice as many
benchmarks and has much higher precision.  In the future, we plan to extend
our framework to unbounded verification through invariant generation.


\bibliographystyle{splncs03}
\bibliography{biblio.bib}

\newpage
\appendix
\section*{Appendix}
\section{Detailed Experimental Results}\label{appendix:extended_result}
Table~\ref{detailed_result} gives a detailed comparison between CBMC version
5.5 and ACDLP.  Columns~1--4 in Table~\ref{detailed_result} contain the
name of the tool, the benchmark category, the number of lines of code (LOC),
and the total number of safe and unsafe benchmarks in the respective
categories (labelled as Safe/Unsafe).  The solver statistics ({\em
Decisions, Propagations, Conflicts, Conflict Literals, Restarts}) 
for CBMC and ACDLP are in columns~5--9.
%

We classify our benchmarks into separate categories. 
We label the benchmarks in bit-vector regression category from SV-COMP'16 
as {\em Bit-vector}, ANSI-C models of hardware circuits auto-generated by v2c 
tool as {\em Verilog-C} and auto-generated Controller code and control-intensive 
benchmarks as {\em Control-Flow} category.  The total number of benchmarks in 
{\em bit-vector} category are 13, {\em Control-Flow} category contains 
55 benchmarks and {\em Verilog-C} category has 17 benchmarks. The timeout for 
our experiments is set to~200 seconds.  All times in Table~\ref{detailed_result} 
and Table~\ref{ai-result} are in seconds. 

The Bit-vector category contains a total of~13 benchmarks, out of which~6
are safe and the remaining~7 are unsafe benchmarks.  The benchmarks in the
control-flow category contains simple bounded loop analysis with relational
properties to more complex controller code containing nested loops with
varying loop bounds.  Out of~55 benchmarks in this category, 35 are safe and
20 are unsafe.  We verified a total of~17 hardware benchmarks, which are
given in Verilog RTL language.  Out of these~17 benchmarks,~10 are safe and
the remaining~7 are unsafe.  The software models (in ANSI-C) for the Verilog
circuits are obtained via a Verilog to C translator tool, {\em v2c}.  These
software models are then fed to CBMC and ACDLP.  The hardware benchmarks
include an implementation of a Instruction buffer logic, FIFO arbiter,
traffic light controller, cache coherence protocol, Dekker's mutual
exclusion algorithm among others.  The largest benchmark is the cache
coherence protocol which consists of~890 LOC and the smallest benchmark is
TicTacToe with~67 LOC.  The software models of these Verilog circuits uses
several complex bit-wise logic to map hardware operations into an equivalent
C syntax.  We emphasize that our implementation can handle bit-wise
operations out-of-the-box.

\begin{table}[!b]
\begin{center}
{
\begin{tabular}{l|l|r|r|r|r|r|r|r}
\hline
           &          &     & Safe/  &           & Propa-  &           & Conflict &          \\
  Verifier & Category & LOC & Unsafe & Decisions & gations & Conflicts & literals & Restarts \\ \hline
  CBMC & \multirow{2}{*}{Bit-vector} & \multirow{2}{*}{501} &
  \multirow{2}{*}{6/7} & 1011 & 1190 & 0 & 0 & 7 \\
  ACDLP & & & & 0 & 44 & 0 & 0 & 0 \\ \hline
  CBMC & \multirow{2}{*}{Control-Flow} & \multirow{2}{*}{1387} & 
  \multirow{2}{*}{35/20} & 29382 & 379727 & 4520 & 37160 & 62 \\ 
  ACDLP & & & & 414 & 6487 & 195 & 180 & 0  \\ \hline
  CBMC & \multirow{2}{*}{Verilog-C} & \multirow{2}{*}{4210} & 
  \multirow{2}{*}{10/7} & 131932 & 322707 & 69 & 349 & 6 \\ 
  ACDLP & & & & 625 & 8196 & 22 & 22 & 0 \\ \hline
\end{tabular}
}
\end{center}
\caption{CBMC versus ACDLP}
\label{detailed_result}
\end{table}

The statistics for ACDLP in Table~\ref{detailed_result} is obtained using an
ordered decision heuristic, multi-way propagation heuristic and a first-UIP
learning heuristic.  Note that the deductions using a multi-way heuristic is
more precise than forward or backward heuristics, but multi-way heuristic
takes longer time to reach the fixed-point.  Furthermore, multi-way
heuristic significantly reduces the total number of decisions, propagations
and learning iterations due to higher precision of the deductions made in
the abstract domain.  Overall, ACDLP reduces the total number of decisions,
propagations, conflicts and restarts by a factor of~20X compared to CBMC.

\begin{table}[t]
\begin{center}
{
\begin{tabular}{l|l|r|r|r}
\hline
  Verifier & Category & \#Proved (safe/unsafe) & \#Inconclusive & \#False Positives \\ \hline
  Astr{\'e}e & \multirow{2}{*}{Bit-vector} & 5/7 & 0 & 1 \\
  ACDLP & & 6/7 & 0 & 0 \\ \hline
  Aste{\'e}e & \multirow{2}{*}{Control-Flow} & 24/9 & 0 & 22 \\
  ACDLP & & 35/17 & 3 & 0 \\ \hline
  Astr{\'e}e & \multirow{2}{*}{Verilog-C} & 2/4 & 0 & 11 \\
  ACDLP & & 9/7 & 1 & 0 \\ \hline
\end{tabular}
}
\end{center}
  \caption{Astr{\'e}e versus ACDLP}
\label{ai-result}
\end{table}

Table~\ref{ai-result} gives a detailed comparison between Astr{\'e}e and
ACDLP.  Columns~1--5 in Table~\ref{ai-result} gives the name of the
tool, the benchmark category, the total number of instances proved safe or
unsafe (labelled as safe/unsafe), the total number of inconclusive
benchmarks and total number of false positives per category.

Table~\ref{ai-result} shows that ACDLP solved twice more benchmarks than
Astr{\'e}e.  The total number of inconclusive results in ACDLP is~4.  The
inconclusive results is because of timeout.  By~contrast, Astr{\'e}e reports
a total of~53 false positives among~85 benchmarks.  Clearly, ACDLP is more
precise than Astr{\'e}e.

\section{Decision Heuristics in ACDLP}~\label{decision}
We have implemented several decision heuristics in ACDLP: {\em ordered}, 
{\em longest-range}, {\em random}, and the {\em activity based} 
decision heuristic.  The {\em ordered} decision heuristic 
makes decisions on meet irreducibles that involve conditional 
variables (variables that appear in conditional branches) first 
before choosing meet irreducibles with numerical variables.  
%
The {\em longest-range} heuristic simply keeps track of the bounds
$\numabsval_l,\numabsval_u$ of matching template rows, which are 
row vectors $\vec{c}$, $\vec{c}'$ such that $\vec{c}=-\vec{c}'$.
$\numabsval_l\leq \vec{c}\vec{x}\leq \numabsval_u$, picks the one with the longest range
$\numabsval_u-\numabsval_l$, and randomly returns the meet irreducible
$\vec{c}\vec{x}\leq
\lfloor\frac{\numabsval_l+\numabsval_u}{2}\rfloor$ or its
complement. This ensures a fairness policy in selecting a variable
since it guarantees that the intervals of meet irreducibles are
uniformly restricted.
The {\em random} decision heuristic arbitrarily picks a meet irreducible  
for making decision. 
%
%
The {\em activity based} decision heuristic is inspired by the 
decision heuristic used in the Berkmin SAT solver.  
The activity based heuristic 
keeps track of the activity of meet irreducibles that 
participate in conflict clauses.  Based on the most 
active meet irreducible, ranges are split similar to 
the {\em longest-range} heuristic.
\section{Decisions, Propagations and Learning in ACDLP}
%

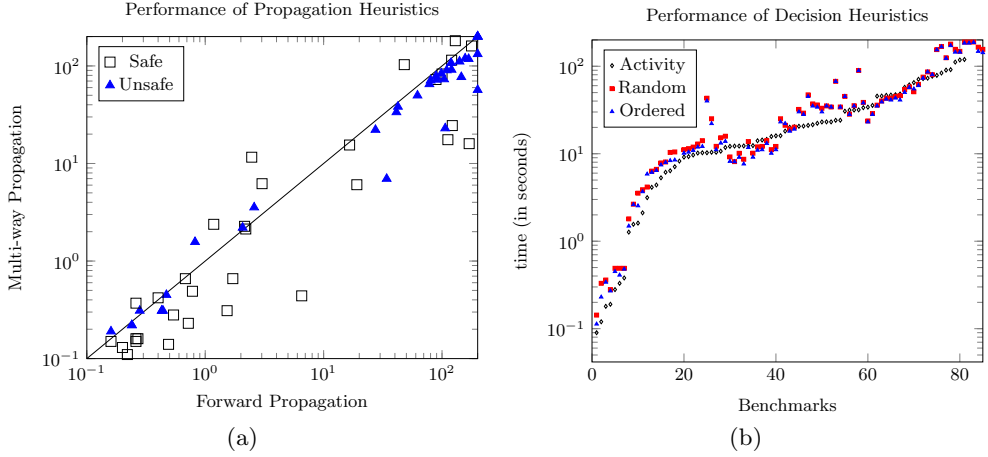
\begin{figure}[t]
\begin{tabular}{@{\hspace{-1.5em}}c@{\hspace{1em}}c}
\begin{tikzpicture}[scale=0.75]
	\begin{loglogaxis} [
      xmin=0.1,xmax=200, ymin=0.1, ymax=200, 
      xlabel=Forward Propagation, ylabel=Multi-way Propagation, 
      title={Performance of Propagation Heuristics},
			legend pos = north west,
			]
\addplot [mark size=1pt,only marks,scatter,point meta=explicit symbolic,
	scatter/classes={s={mark=square,mark size=2.5},u={mark=triangle*,blue,mark size=2.5}}] 
	table [meta=label] {plotdata/scatter-chaotic-forward.dat};
	\legend{Safe,Unsafe}
  \addplot [domain=.1:200] {x};
\end{loglogaxis}
\end{tikzpicture}
 &
\begin{tikzpicture}[scale=0.75]
\pgfplotscreateplotcyclelist{markstyles}{%
mark=diamond\\%
mark=square*,red\\%
mark=triangle*,blue\\%
}
	\begin{axis} [
  title={Performance of Decision Heuristics}, ymode = log,
  xmin=0, xmax=85, only marks,
  ymin=0, ymax=200, 
  xlabel={Benchmarks},
  ylabel={time (in seconds)},
  xtick={},
  ytick={}, 
	legend pos=north west, mark size=1pt, cycle list name=markstyles,
 ]
 \addplot table[x=benchmark,y=time] {plotdata/berkmin.dat};
 \addplot table[x=benchmark,y=time] {plotdata/random.dat};
 \addplot table[x=benchmark,y=time] {plotdata/ordered.dat};
 \addlegendentry{Activity}
 \addlegendentry{Random}
 \addlegendentry{Ordered}
 \end{axis}
\end{tikzpicture} \\
(a) & (b)
\end{tabular}
\caption{\label{prop-dec}
Effect of Propagation Heuristics and Decision Heuristics in ACDLP
}
\end{figure}

%
\paragraph{Propagation Strategy.}
Fig.~\ref{prop-dec}(a) presents a comparison between the {\em forward} 
and {\em multi-way} propagation strategy in ACDLP.  The
choice of strategy influences the total number of decisions and clause 
learning iterations.  Hence, the propagation strategy has a
significant influence on the runtime, which can be seen in
Fig.~\ref{prop-dec}(a).  We did not report the performance 
of {\em backward} propagation strategy due to large 
number of timeouts.  Compared to forward propagation, the multi-way
strategy may take more iterations to reach the fixed-point, but it
subsequently reduces the total number of decisions and conflicts to prove the
program.  This is attributed to the higher precision of the meet irreducibles 
inferred by the multi-way strategy, which subsequently aids the decision 
heuristics to make better decisions.  

\paragraph{Decision Heuristics.}
Fig.~\ref{prop-dec}(b) shows the performance of different decision
heuristics in ACDLP.  Note that the runtimes for all decision heuristics are
obtained using the multi-way propagation strategy.  The runtimes are very
close, but we can still discern some key characteristics of these
heuristics.  The activity based heuristic performs consistently well for most safe
benchmarks and all bit-vector category benchmarks.  By contrast, the ordered
heuristic performs better for programs with conditional branches since it
prioritises decisions on meet irreducibles that appear in conditionals.  The
runtimes for the random heuristic are marginally higher than the other
two.  This suggests that domain-specific decision heuristics are important
for ACDLP.
\Omit{
Whereas activity-based heuristics such as Berkmin heuristic which 
works well in propositional cases performs best for benchmarks 
that encountered the maximum number of conflicts to prove safety, 
thus allowing the heuristics to choose the decision variable among the set of learnt clauses.   
}

\paragraph{Learning.}
Learning has a significant influence on the runtime of ACDLP.  We~compare
the UIP-based learning technique with an analysis that performs classical 
DPLL-style analysis.
The effect of UIP computation allows ACDLP to backtrack non-chronologically 
and guide the model search with a learnt transformer.  But classical 
DPLL-style analysis exhibits case-enumeration behaviour and could not finish 
within the time bound for 20\% of our benchmarks.
\section{Computing Lazy Closure for Template Polyhedra}\label{appendix:lazyclosure}
Computing the closure for relational domains, such as octagons, is
expensive.  An~advantage of our formalism in
Eq.~(\ref{eq:at2}) is that the \emph{closure} operation for relational
domains can be computed in a lazy manner through the construction of a
subdomain.  A subdomain $\subdomain$ is constructed from domain $\domain$
for a abstract deduction transformer $\abstransel{\subdomain}$, such that
$\subdomain=\makesubdomain_\domain(\subvars)$, where $\subvars$ are
variables that appears in $\abstransel{\subdomain}$.  The construction of
$\subdomain$ allows us to perform one step of the closure operation when
$\abstransel{\subdomain}$ is applied.
For example, let us consider $\domain=\octagons[\{x,y,z\}]$ and
$\subvars=\{y\}$. An octagonal inequality relates at 
most two variables. Thus it is sufficient to consider the subdomain
$\makesubdomain_\domain(\{y\})=\octagons[\{y\}]\cup\octagons[\{x,y\}]\cup\octagons[\{y,z\}]$,
which will compute the one-step transitive relations of~$y$ with each
of the other variables. 
Only if a abstract deduction transformer subsequently makes new deductions 
on $x$ or $z$, then the next step of the closure will be computed through 
the subdomain $\octagons[\{x,z\}]$.
\Omit{
Hence, when the abstract deduction transformer is applied we do not
compute the full closure in the full domain,
but we compute only a single step of the closure in a restricted
domain, which makes single deduction steps more efficient.
}
Hence, an application of abstract deduction transformer does not 
compute the full closure in the full domain, but compute only a 
single step of the closure in a restricted domain. This makes 
each deduction step more efficient but may require more steps 
to reach the fixed point.

Let us demonstrate the idea of lazy closure with a concrete example.  Assume
the program in the left of Figure~\ref{fig:lazy}.  The corresponding
locations are marked L1 and L2.  We~analyze the program with an octagon
domain ($\octagons$), which computes the closure in lazy manner.  The lazy
closure computation in octagon domain is shown in the right of
Figure~\ref{fig:lazy}.

Recall that a closure in octagon domain achieves a normal form by computing all
implied constraints among numerical variables. The closure operation is 
necessary to perform precise domain operations.  The ACDLP analysis in
Figure~\ref{fig:lazy} performs forward propagation in $\octagons$ by creating a 
subdomain $\subdomain$ for every transformer using the function $\makesubdomain$.  
Note that the choice of subdomain over lhs variables of the transformers guides the 
analysis in forward direction in this example. The subdomain 
corresponding to L1 over $y$ is given by 
$\octagons[\{y\}]\cup\octagons[\{y,z\}]$.  This means, only those deductions 
which are implied by the domain $\octagons[\{y\}]\cup\octagons[\{y,z\}]$ can be
inferred at L1.  No deductions over $\octagons[\{y,x\}]$ is performed at L1.  
Thus, we delay the deductions over $\{y,x\}$ until we encounter an abstract 
transformer over these variables.  This does not admit a normal form for 
octagonal constraints after the application of the transformer at L1, but it 
makes the deduction step at L1 more efficient.  

Assume that the initial abstract value $(\absval)$ is $\absval=(x=y)$. 
Then, the deduction at L1 infers ${y=z}$.  Thus, the updated abstract value
is $\absval=\{x=y \wedge y=z\}$.  We now analyze the transformer at L2.  The
subdomain for L2 over variable $x$ (for forward propagation) is given by
$\octagons[\{x\}]\cup\octagons[\{x,y\}]\cup\octagons[\{x,z\}]\cup\octagons[\{x,w\}]$. 
Note that we delayed the deduction over $\octagons[\{x,y\}]$ at L1, but only
perform the deductions over $\octagons[\{x,y\}]$ at L2.  This is the notion
of lazy closure computation.  The new deductions at L2 are $\{x=z, x-w \leq
1, x-w \geq 1\}$ and the final abstract value is $\absval=\{x=y
\wedge\allowbreak y=z \wedge\allowbreak x=z \wedge\allowbreak x-w \leq 1
\wedge x-w \geq 1\}$.  Thus, the normal form over $\octagons{\{x,y,z\}}$ is
only achieved at L2.  However, we do not perform deductions over
$\octagons{\{w,z\}}$ at L2, which is delayed until the point where we
encounter an abstract transformer that forces us to infer such deductions.
\begin{figure}[htbp]
\centering
\begin{tabular}{c|c}
\hline
C program & Lazy Closure Computation \\
\hline
\scriptsize
\begin{lstlisting}[mathescape=true,language=C]
int main() {
  L1: y=z;
  L2: x=w+1;
}
\end{lstlisting}
&
\begin{lstlisting}[mathescape=true,language=C]
L1: $\makesubdomain_\domain(\{y\})=\octagons[\{y\}]\cup\octagons[\{y,z\}]$
    New Deductions: $\{y=z\}$
    Abstract Value: $\absval=\{x=y \wedge y=z\}$
L2: $\makesubdomain_\domain(\{x\})=\octagons[\{x\}]\cup\octagons[\{x,y\}]\cup\octagons[\{x,z\}]\cup\octagons[\{x,w\}]$
    New Deductions: $\{x=z, x-w \leq 1, x-w \geq 1\}$
    Abstract Value: $\absval=\{x=y \wedge y=z \wedge x=z \wedge x-w \leq 1 \wedge x-w \geq 1\}$
\end{lstlisting}
\\
\hline
\end{tabular}
\caption{\label{fig:lazy}
C Program and Lazy closure operation for Octagons}
\end{figure}

\section{Example Demonstrating Execution of ACDLP}
\begin{figure}
\scriptsize
\centering
\begin{tabular}{c}
\hline
  Control-Flow Graph \\
\hline
\begin{minipage}{4.2cm}
\scalebox{.65}{\import{figures/}{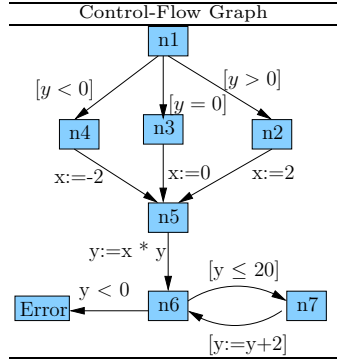}}
\end{minipage}
\\ 
\hline
\end{tabular}
\caption{\label{fig:se} A Control-Flow Graph of a Program}
\end{figure}

\begin{figure}[t]
\centering
\vspace*{-0.2cm}
\scalebox{.60}{\import{figures/}{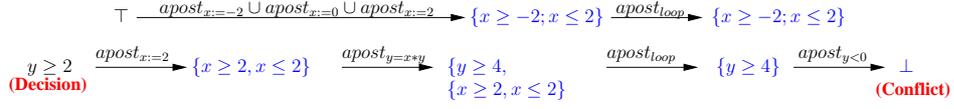}}
  \caption{Model Search as Downward Iteration Sequence with Decisions and
  Deductions}
\label{model-search}
\end{figure}

\begin{figure}
\centering
\vspace*{-0.2cm}
\scalebox{.60}{\import{figures/}{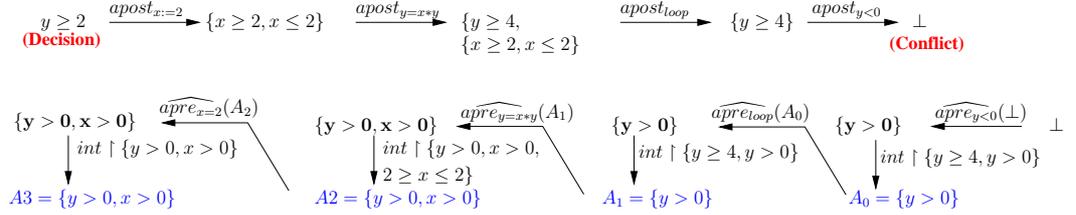}}
  \caption{Conflict Analysis with underapproximate weakest precondition and
  heuristic generalization}
\label{conflict-example}
\end{figure}

We now present a step-by-step execution of the ACDLP algorithm. 
Figure~\ref{fig:se} shows a Control-flow Graph of a program. The program is
\emph{safe} since the \texttt{Error} location is unreachable along every
execution of program paths.  We analyze the safety of this program using ACDLP.
Fig.~\ref{model-search} shows an example run for the counterexample search procedure 
for the program in~Fig.\ref{fig:se}.  The elements obtained using an overapproximate 
strongest postcondition transformer $apost$ are marked in blue in Fig.~\ref{model-search}.  
Starting from $\top$, forward analysis concludes that $x$ is between -2 and 2
from $apost_{y:=-2} \cup apost_{y=0} \cup apost{y:=2}$.  Note that the loop is completely 
unwound and all statements corresponding to the loop are collectively referred to as $loop$. 
A forward fixed-point analysis (marked by $apost_{loop}$) does not yield any new 
information. Clearly, the analysis
is not precise to infer anything about the reachability of the error location $Error$. 
Hence, we apply a decision by picking a meet irreducible $y\geq 2$ to increase the 
precision of analysis.  We then apply forward analysis from this decision which
yields a downward iteration sequence as shown in lower part of
Fig.~\ref{model-search}.  Forward analysis concludes that $\{y \geq 4\}$.
This leads to \emph{conflict}, marked as $\bot$. Hence, the error location
$Error$ is unreachable for this decision. 
%
The conflict analysis procedure is
shown in Figure~\ref{conflict-example}.  We iteratively apply the weakest
precondition transformer $\widehat{apre}$ 
starting from the conflict element ($\bot$), the result of which is shown in
bold text.  For example, $\widehat{apre_{y \leq 0}}(\bot)= \{y>0\}$; whereas the
result of strongest postcondition is 
is $\{y\geq 4\}$.  So, we heuristically pick a generalized element $a$ such 
 that $\{y\geq 4\} \sqsubseteq a \sqsubseteq \{y>0\}$; 
  we pick $c=\{y>0\}$ through the application of heuristic choice~\cite{leo-thesis} 
  (corresponds to upwards interpolation), $int\upharpoonright(y>0, y\geq4)$, marked in blue.  
  Note that the loop is completely unwound and all statements corresponding 
  to the loop are collectively referred to as $loop$.
We then repeat the process marked by $\widehat{apre_{loop}}$. 
Subsequently, we derive a generalized reason, $\{x>0, y>0\}$, that strictly generalizes
the decision $x>=2$. The conflict reason $\{x>0,y>0\}$ underapproximates 
set of safe traces. \\
\begin{figure}[htbp]
\centering
\vspace*{-0.2cm}
\scalebox{.60}{\import{figures/}{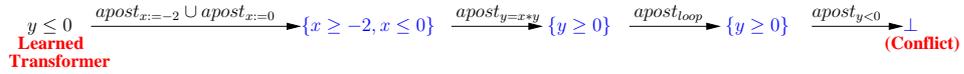}}
\caption{Downward Iteration Sequence with Learned transformer \label{learning}}
\end{figure}
Fig.~\ref{learning} shows the sequence of fixed point iteration with the learned
transformer $y \leq 0$, obtained from $AUnit$.  Clearly, this also leads to 
\emph{conflict}. There are no further cases to explore. Thus, the procedure 
terminates and returns \emph{safe}.

\end{document}